\documentclass[3p,onecolumn]{elsarticle}

\usepackage{graphicx}
\usepackage{dcolumn}
\usepackage{pifont}
\usepackage{bm}
\usepackage{multirow}
\usepackage{amsmath}
\usepackage{float}
\usepackage{txfonts}
\usepackage[version=3]{mhchem}

\journal{Nano Energy}

\begin{document}
\title{{\it Ab initio} study of sodium cointercalation with diglyme molecule into graphite}

\author[kimuniv-m]{Chol-Jun Yu\corref{cor}}
\ead{ryongnam14@yahoo.com}
\author[kimuniv-m]{Song-Bok Ri}
\author[kimuniv-m]{Song-Hyok Choe}
\author[kimuniv-m]{Gum-Chol Ri}
\author[kimuniv-m]{Yun-Hyok Kye}
\author[kimuniv-m]{Sung-Chol Kim}

\cortext[cor]{Corresponding author}

\address[kimuniv-m]{Department of Computational Materials Design, Faculty of Materials Science, Kim Il Sung University, \\ Ryongnam-Dong, Taesong District, Pyongyang, Democratic People's Republic of Korea}

\begin{abstract}
The cointercalation of sodium with the solvent organic molecule into graphite can resolve difficulty of forming the stage-I Na-graphite intercalation compound, which is a predominant anode of Na-ion battery. To clarify the mechanism of such cointercalation, we investigate the atomistic structure, energetics, electrochemical properties, ion and electron conductance, and charge transferring upon de/intercalation of the solvated Na-diglyme ion into graphite with {\it ab initio} calculations. It is found that the Na(digl)$_2$C$_n$ compound has the negatively lowest intercalation energy at $n\approx21$, the solvated Na(digl)$_2$ ion diffuses fast in the interlayer space, and their electronic conductance can be enhanced compared to graphite. The calculations reveal that the diglyme molecules as well as Na atom donates electrons to the graphene layer, resulting in the formation of ionic bonding between the graphene layer and the moiety of diglyme molecule. This work will contribute to the development of innovative anode materials for alkali-ion battery applications.
\end{abstract}


\maketitle

Energy and environmental issues become increasingly global, actual and vital to all the nations on the earth. This prompts people to cease from mass consumption of fossil fuels, which are being exhausted and moreover cause greenhouse gas emission, and to create new technologies for utilizing renewable and clean energy sources such as solar and wind power. However, these sources are changeable in time and widespread in space. For solar and wind power to become competitive with fossil fuels in the electricity market, therefore, it is necessary to develop an efficient energy storage system as well as an effective energy harvesting device. Although Li-ion batteries (LIBs) have been widely used for portable electronics during the past two decades due to their high power density, growing concerns for low abundance and high price of lithium resource~\cite{Tarascon01,Tarascon10,Goodenough10} triggered extensive research on Na-ion batterries (NIBs) recently considering high abundance and low cost of sodium precursor~\cite{Palomares,Slater,Yabuuchi14,Sawicki}. When compared to Li$^+$ ion, however, Na$^+$ ion has lower ionization potential and larger ionic radius, which make it difficult to find suitable anode materials~\cite{Islam,Wang15}.

Requirements for a good anode material can be (i) high specific capacity and low redox potential for high energy density, (ii) high abundance, low cost and facile fabrication for economics, (iii) low chemical reactivity with electrolyte and small volume change during operation for safety and long cycling life, and (iv) non-toxicity for environmental friendliness. When surveying NIB anodes based on these requirements, metal sodium can be used surely but has a critical problem such as high reactivity with organic electrolyte solvents, while metal oxides and alloy based materials have in general high specific capacities but suffer from large volume change~\cite{Kim14,Wu,Jiang,Mortazavi15}. In spite of relatively low capacity, carbon based materials might be the most widely used anode for NIBs because of good electronic conductivity and excellent electrochemical performance~\cite{Luo,Balogun}. The research tended to mainly non-graphitic carbons like disordered hard carbon, which has high specific capacity over $\sim$300 mAh g$^{-1}$ but still shortcomings such as low coulombic efficiency in the first cycle due to large surface area~\cite{Irisarri,Lihard}. During the last few years, the significant progress in using graphitic carbons has been made~\cite{Li,Ramos,Wen}. From the studies of hard carbons, it was awoken that desirable carbon materials should have long-range order, high interlayer distance, low porosity and small surface area~\cite{Ramos}. In this regard, Wen {\it et al.}~\cite{Wen} suggested the expanded graphite with a mean interlayer distance of $\sim$0.43 nm (0.334 nm in graphite) and a relatively small surface area of $\sim$30 m$^2$ g$^{-1}$, being able to deliver high specific capacity of $\sim$300 mAh g$^{-1}$ at low current density of 20 mA g$^{-1}$. At high current density, however, the modest performance was observed. Using the expanded synthetic graphite materials, Ramos {\it et al.}~\cite{Ramos} increased the current density up to 100 mA g$^{-1}$, at which the capacity was of $\sim$110 mAh g$^{-1}$.

Then, why not graphite itself for NIB anode? Although the alkali metals are known to reversibly intercalate into graphite, forming binary graphite intercalation compounds ($b$-GICs) in first stage, only sodium can form $b$-GICs in high stages~\cite{Dresselhaus,Stevens}. This leads to a very low specific capacity for cells, {\it e.g.}, $\sim$35 mAh g$^{-1}$ for \ce{NaC64} versus $\sim$372 mAh g$^{-1}$ for \ce{LiC6}, hindering practical use of graphite as NIB anode. The reason might be the small interlayer spacing of graphite, {\it i.e.,} big mismatch between graphite interlayer distance and Na ion size, and most likely, weak binding of Na ion (electron donor) to graphene sheet (acceptor) through $\pi$ interaction~\cite{Nobuhara,Wang,Okamoto}. A fortunate side for electrode performance is that the larger radius ions diffuse in graphite more smoothly~\cite{Nobuhara}.

To form stage-I Na-GICs, cointercalation process, {\it i.e.,} intercalation of sodium with solvent organic molecule, has been suggested~\cite{Jache,Kim15,Kim152,Zhu,Maluangnont3, Maluangnont1}. This produces donor-type ternary GICs ($t$-GICs), denoting Na$^+_x$(solv)$_y$C$^-_n$ with C$_n$ the graphene sheet containing $n$ carbon atoms and \ce{(solv)_{\it y}} the $y$ solvent organic molecules~\cite{Jache}. The solvents are typically a series of alkylamines~\cite{Maluangnont1,Maluangnont2,Maluangnont3}, and ether-based electrolytes~\cite{Jache,Kim15,Kim152,Zhu}. Recently, Jache and Adelhelm~\cite{Jache} made use of $t$-GICs containing sodium atom and diglyme (diethylene glycol dimethyl ether: \ce{C6H14O3}) molecule with an estimated stoichiometry of \ce{Na(digl)2C20} as an effective NIB anode, which delivers moderately high reversible capacities close to 100 mAh g$^{-1}$ for 1000 cycles and coulomb efficiencies over 99.87\%. Hasa {\it et al.}~\cite{Hasa} demonstrated the NIB full cell consisted of a graphite anode, glyme-based electrolyte (tetraethylene glycol dimethyl ether (TEGDME)-\ce{NaClO4}), and a layered oxide cathode, {\it i.e.,} Graphite/TEGDME-\ce{NaClO4}/\ce{Na_{0.7}CoO2}, which offers unique characteristics of long cycle life, high efficiency and high power density, though the lower energy density than LIBs.

In spite of such successful applications, there are still concerns about using $t$-GICs as NIB anode~\cite{Balogun}. This is in part due to a lack of sufficient understanding of sodium cointercalation phenomena. {\it Ab initio} calculations within the framework of density functional theory (DFT) have proven to give valuable insights into electrochemical performance of NIB anodes~\cite{Buldum,Mortazavi,Tsai,yucj06}. In this work, we perform systematic {\it ab initio} DFT calculations of $t$-GICs cointercalated with sodium atom and diglyme molecule, aiming to clarify the mechanism of formation and operation as NIB anodes.

\section{\label{sec:theoretics}Computational methods}
\begin{figure*}[!t]
\centering
\includegraphics[clip=true,scale=0.41]{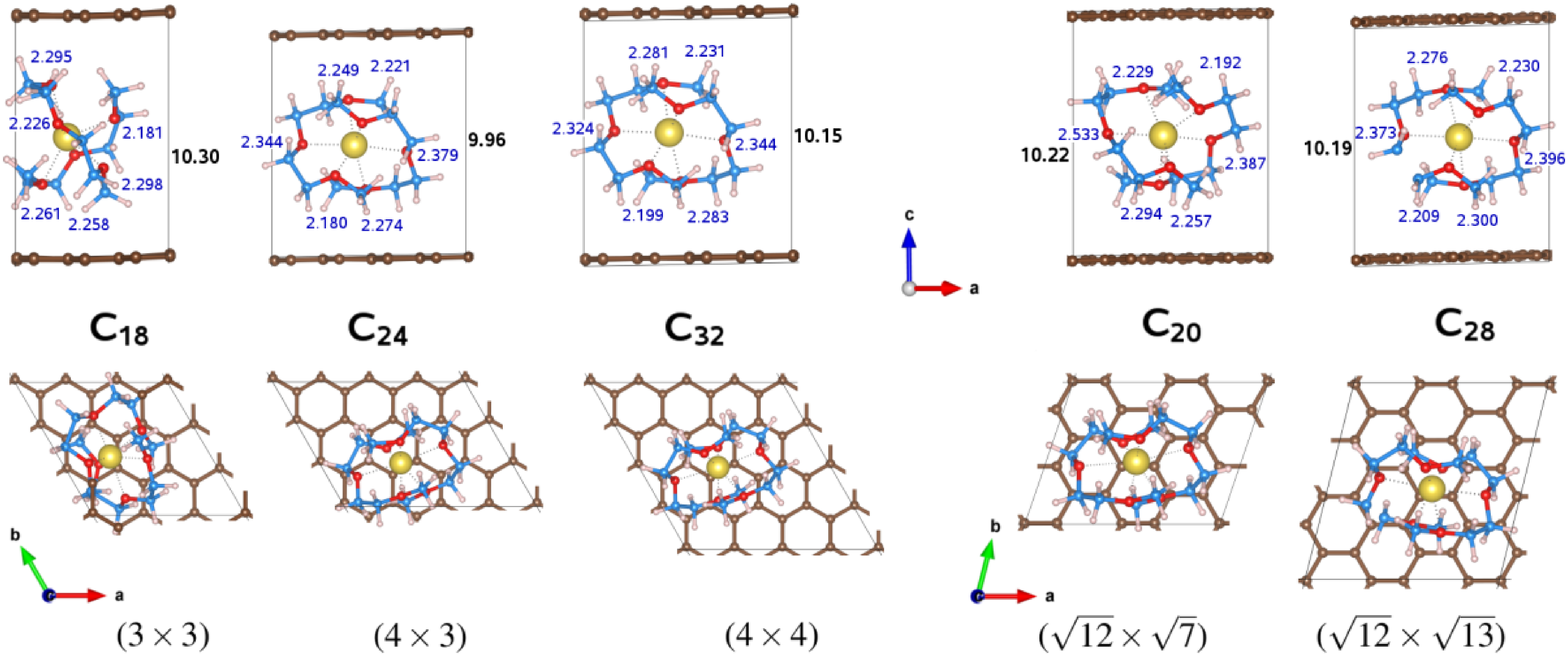}
\caption{Side (top panel) and top view (bottom panel) of unit cells of Na(digl)$_2$C$_n$ GICs with $n=$18, 24, 32, 20, and 28, having ($3\times3$), ($4\times3$), ($4\times4$), ($\sqrt{12}\times\sqrt{7}$), and ($\sqrt{12}\times\sqrt{13}$) cells with $AA$-stacked graphene layers. Carbon atoms of graphene sheet are represented by brown balls (hexagon), while carbon, oxygen and hydrogen atoms of diglyme molecule by small blue, red and white balls, and sodium atoms by big yellow balls. Interlayer distance and six Na$-$O bond lengths in \AA~unit optimized by PBEsol+vdW method are indicated in top panel.}
\label{fig_model}
\end{figure*}

Firstly we make models for the crystalline $t$-GICs cointercalated with sodium atom and diglyme molecule, Na$_x$(digl)$_y$C$_n$ with $x=1$, $y=2$, and different $n$ numbers. Regarding the intercalant complexes, it was assumed in the previous works~\cite{Jache,Henderson,Rhodes,Matsui} that two diglyme molecules are bound to one alkali ion with six coordination number by ether oxygen atoms from the solvent molecules, and thus we make a model of solvated ion as one sodium atom surrounded by two crossing diglyme molecules, {\it i.e.}, \ce{Na(digl)2} (for reference, one Na and one diglyme \ce{Na(digl)1}, and one Li and two diglymes \ce{Li(digl)2} are also considered). Note that different solvated ion models with two alkali atoms and two diglyme molecules \ce{Na2(digl)2} were also suggested and studied in other works~\cite{Kim15,Kim152,Zhu}. Regarding the graphite hosts with different carbon numbers $n$, we use ($3\times3$), ($4\times3$), ($4\times4$), ($\sqrt{12}\times\sqrt{7}$), and ($\sqrt{12}\times\sqrt{13}$) cells, which correspond to $n=18$, 24, 32, 20, and 28, respectively. The solvated ions are assumed to be intercalated into $AA$-stacked graphene layers, as shown in Fig.~\ref{fig_model}.

The {\it ab initio} DFT calculations were carried out using the pseudopotential-pseudo atomic orbital (PAO) method as implemented in the SIESTA package~\cite{SIESTA}. We have tested different exchange-correlation (XC) functionals, such as the Perdew-Zunger formalism within local density approximation (LDA)~\cite{PZlda}, and the Perdew-Burke-Ernzerhof (PBE)~\cite{PBE}, the PBE for solid (PBEsol)~\cite{PBEsol} and the Lee-Yang-Parr (LYP)~\cite{LYP} functionals within generalized gradient approximation (GGA). It can be thought that the dispersive van der Waals (vdW) interactions between graphene sheets may play an important role in Na(digl)-cointercalated graphite as well as pristine graphite~\cite{Chen, Fernandez, yucj04}. To estimate such vdW interactions, we used the semi-empirical Grimme's approach~\cite{DFTD2}, where the vdW energy is expressed as follows,
\begin{equation}
E_\text{vdW}=-s_6\sum_{j<i}\frac{C_6^{ij}}{r_{ij}^6}\frac{1}{1+e^{-d(r_{ij}/R_{ij}^{0}-1)}}
\end{equation}
where $r_{ij}=\vert \textbf{r}_i-\textbf{r}_j \vert$ the distance between the $i$-th and $j$-th atoms, $C_6^{ij}=\sqrt{C_6^iC_6^j}$ being $C_6$ the dispersion coefficient of atom, $R_{ij}^0=R_i^0+R_j^0$ being $R^0$ the vdW radius of atom, and $s_6$ and $d$ are the scaling and damping parameters, respectively. We took the values of $C_6$ and $R^0$ for all atoms from Ref.~\cite{DFTD2} and used the default values of 1.66 for $s_6$ and 20.0 for $d$ given in the SIESTA code. Note that some papers reported negligible vdW effect for $t$-GICs~\cite{yucj06,Chen}, and the localized basis sets like PAO might not properly describe the vdW force, compared to expanded basis sets like plane wave~\cite{yucj09}.

For all the atoms, standard double-$\zeta$ polarized (DZP) basis sets were generated with 300 meV energy shift for orbital-confining cutoff radii and 0.25 split norm for the split-valence of basis. We constructed the Troullier-Martins~\cite{TMpseudo} norm-conserving pseudopotentials of atoms using the ATOM code provided in the package, where the valence electron configurations are Na-3s$^1$3p$^0$3d$^0$4fs$^0$, Li-2s$^1$2p$^0$3d$^0$4fs$^0$, C-2s$^2$2p$^2$3d$^0$4fs$^0$, O-2s$^2$2p$^4$3d$^0$4fs$^0$, H-1s$^1$2p$^0$3d$^0$4fs$^0$. The non-linear core correction was applied to Na with a core radius of 1.50 Bohr and Li with 0.7 Bohr. The transferability testings were performed for all the generated pseudopotentials. Spin polarization was considered. We have conducted the convergence test of total energy with respect to the major computational parameters$-$plane-wave cutoff energy and $k$-grid cutoff length. After that, they were set to be 300 Ry and 10 \AA, respectively, which guarantee an accuracy of total energy of 5 meV per cell (Fig.S1\dag). The crystalline lattice parameters and internal atomic coordinates were fully relaxed until the forces on each atom converged to within 0.02 eV \AA$^{-1}$ and the stress on lattice to within 0.01 GPa.

To estimate the possibility of compound formation, we calculate the total energy difference between the compound and its reactants, as the entropic and enthalpic contributions might be negligible. First, upon the formation reaction of solvated sodium or lithium A$_x$(digl)$_y$ ($\text{A}=\text{Na}$ or Li, $x, y=1$ or 2) complexes,
\begin{equation}
\label{eq_solv}
x\text{A}+y(\text{digl})\rightarrow \text{A}_x(\text{digl})_y
\end{equation}
the binding energy can be calculated using the total energies of isolated A$_x$(digl)$_y$ complex, diglyme molecule and alkali atom,
\begin{equation}
\label{eq_Ebind}
E_\text{b}=E_{\text{A}_x(\text{digl})_y}-(xE_\text{A}+yE_\text{digl})
\end{equation}
The cubic supercells with a lattice constant of 40 \AA~were used to model the isolated subjects. Then, upon the intercalation of the solvation complex into graphite to form A$_x$(digl)$_y$C$_n$ $t$-GICs,
\begin{equation}
\label{eq_tgic}
\text{A}_x(\text{digl})_y+\text{C}_n\rightarrow \text{A}_x(\text{digl})_y\text{C}_n,
\end{equation}
the intercalation energy can be calculated as follows~\cite{Tasaki},
\begin{equation}
\label{eq_Eint}
E_\text{int}=E_{\text{A}_x(\text{digl})_y\text{C}_n}-(E_{\text{A}_x(\text{digl})_y}+E_{\text{C}_n}).
\end{equation}
The negative binding or intercalation energy indicates that the formation of compound is , {\it i.e.}, spontaneous. The strength of binding between graphene sheets in graphite and GICs can be estimated by the exfoliation energy per carbon atom,
\begin{equation}
\label{eq_Eexf}
E_\text{exf}=\frac{1}{n_\text{C}}[E(d_\text{i}=d_\text{e})-E(d_\text{i}=\infty)],
\end{equation}
where $n_\text{C}$ is the number of carbon atoms in graphene sheet, $d_\text{e}$ the equilibrium interlayer distance and $E(d_\text{i}=\infty)$ was replaced by $E(30$~\AA$)$ in this work due to little change in the energy beyond $d_\text{i}=30$~\AA. In these energetic estimations, basis set superposition error (BSSE) was not corrected, because it was reported that the GICs with a strong binding between the intercalant and the graphene sheet are not proper to apply the BSSE correction method~\cite{Tasaki}.

\section{\label{sec-result}Results and discussion}
\subsection{\label{subsec-xc}Assessment of XC functional and vdW inclusion}
It was in general accepted that the lattice-related and energetic properties of graphite and GICs depend sensitively on the choice of XC functional and whether the inclusion of vdW correction or not, as evidenced in our previous works~\cite{yucj04,yucj06}. Therefore, it is necessary to evaluate the reliability of various XC functionals and vdW inclusion. We computed the energetic quantities such as binding, intercalation, and exfoliation energies, which are indicative factors whether the compound can be formed spontaneously or not, employing one LDA functional (PZ) and three GGA functionals (PBE, PBEsol and LYP) with and without vdW correction.

\begin{figure}[!b]
\centering
\includegraphics[clip=true,scale=0.23]{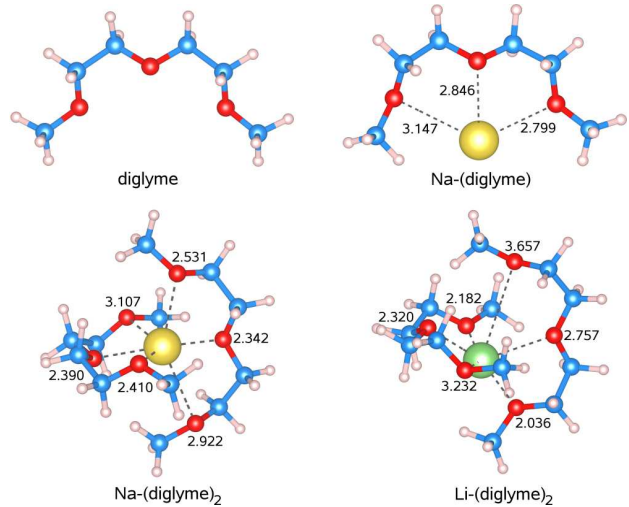}
\caption{Molecular structures of free diglyme molecule and Na-(diglyme), Na-(diglyme)$_2$ and Li-(diglyme)$_2$ complexes optimized by PBEsol+vdW method. Carbon, oxygen and hydrogen atoms of diglyme molecule are represented by small blue, red and white balls, and sodium and lithium atoms by big yellow and green balls. A$-$O bond lengths are indicated.}
\label{fig_freemol}
\end{figure}
We begin with an analysis of binding energy calculation of free A$_x$(digl)$_y$ ($\text{A}=\text{Na}$ or Li, $x, y=1$ or 2) complexes. As it was confirmed from experiments that the alkali ions in the crystalline phases of A-glyme compounds are coordinated by six oxygen atoms~\cite{Jache,Henderson,Rhodes}, two diglyme molecules per alkali atom were suggested to reach a stable solvation shell in dilute solution~\cite{Jache}. Moreover, it was shown from theoretical studies on diglyme-solvated alkali ions that the solvation shell still consists of two diglyme molecules in the vicinity of the electrode surface~\cite{Matsui}. Therefore, we considered mainly Na(digl)$_2$ complex in this work and additionally Na(digl) and Li(digl)$_2$ complexes for reference, although Na$_2$(digl)$_2$ complexes are not ruled out~\cite{Kim152}. Fig.~\ref{fig_freemol} shows the molecular structures of diglyme molecule and these solvation complexes optimized by PBEsol+vdW method. Note that the initial configurations for geometrical optimizations were selected from conformation searching with the systematic grid scan method. While Na and Li binding energies in Na(digl) and Li(digl)$_2$ are negative from all XC functionals with and without vdW correction as listed in Table~\ref{tab_xcene}, it was observed that the binding energy of Na to two diglyme molecules in Na(digl)$_2$ are negative only from LDA, LDA+vdW and PBEsol+vdW functionals. When including the vdW correction, the binding energies become slightly lower than those without vdW energy addition. Regarding the cohesive energy ($E_\text{coh}$) of Na and Li metals, PBE and PBEsol could give the closest values to the experiment, with a slightly lowering by including the vdW interaction.
\begin{table*}[!t]
\caption{Lattice constants ($a$ and $c$) and exfoliation energy ($E_\text{exf}$) in graphite, cohesive energy in sodium and lithium metals ($E_\text{coh}$), binding energy in free solvation complex A(digl)$_x$ ($E_\text{b}$), intercalation energy in A(digl)$_x$C$_{24}$ ($E_\text{int}$), exfoliation energy in A(digl)$_x$C$_{20}$, and average electrode voltage ($V_\text{el}$)}
\label{tab_xcene}
\begin{tabular}{lcccccccccc}
\hline
& & LDA & LDA  & PBE & PBE  & PBEsol & PBEsol & LYP & LYP  & Exp. \\
& &     & +vdW &     & +vdW &        & +vdW   &     & +vdW &  \\
\hline
\multirow{3}{*}{Graphite}& $a$ (\AA)&2.466&2.465&2.465&2.464&2.464&2.464&2.471&2.470&2.461$^a$\\
 &$c$ (\AA) &6.217&6.086&6.639&6.509&6.378&6.247&6.816&6.730&6.705$^a$ \\
 &$E_\text{exf}$ (meV)&$-110$&$-130$&$-60$&$-76$&$-77$&$-95$&$-47$&$-61$&$-35\sim-52^b$ \\
\hline
\multirow{2}{*}{$E_\text{coh}$ (eV)}& Na &$-1.31$&$-1.33$&$-1.13$&$-1.16$&$-1.22$&$-1.25$&$-0.91$&$-0.93$ & $-1.11^c$ \\
 & Li &$-1.72$&$-1.74$&$-1.61$&$-1.62$&$-1.64$&$-1.65$&$-1.28$&$-1.30$&$-1.63^c$ \\
\hline
\multirow{3}{*}{$E_\text{b}$ (eV)}& Na(digl) &$-0.42$&$-0.46$&$-0.29$&$-0.32$&$-0.29$&$-0.32$&$-0.28$&$-0.31$ & \\
 & Na(digl)$_2$ &$-0.65$&$-0.80$&~~~0.27&~~~0.12&~~~0.05&$-0.11$&~~~0.27&~~~0.14 & \\
 & Li(digl)$_2$ &$-1.89$&$-2.02$&$-2.20$&$-2.32$&~~~0.00&$-0.13$&$-0.91$&$-1.03$& \\
\hline
\multirow{3}{*}{$E_\text{int}$ (eV)}& Na(digl)C$_{20}$ &$-$2.85&$-$3.06&$-$1.68&$-$1.84&$-$1.83&$-$2.16&$-$1.18&$-$1.40&\\
 & Na(digl)$_2$C$_{20}$ &$-$1.04&$-$1.36&~~~1.01&~~~0.61&~~~0.38&$-$0.02&~~~1.87&~~~1.44& \\
 & Li(digl)$_2$C$_{20}$ &$-3$.11&$-$3.36&$-$0.70&$-$1.18&$-$1.60&$-$1.87&$-$0.12&$-$0.36& \\
\hline
\multirow{3}{*}{$E_\text{exf}$ (meV)}&Na(digl)C$_{20}$ &$-276$&$-330$&$-171$&$-227$&$-226$&$-265$&$-168$&$-206$& \\
 & Na(digl)$_2$C$_{20}$ &$-287$&$-315$&$-202$&$-227$&$-230$&$-255$&$-163$&$-182$& \\
 & Li(digl)$_2$C$_{20}$ &$-283$&$-289$&$-211$&$-232$&$-241$&$-259$&$-180$&$-192$& \\
\hline
\multirow{3}{*}{$V_\text{el}$ (V)}& Na(digl)C$_{24}$ &4.01&4.28&~~~1.23&~~~4.77&~~~2.00&2.00&~~~0.40&~~~1.17& \\
 & Na(digl)$_2$C$_{24}$ &2.70&3.80&$-2.51$&$-1.43$&$-0.93$&0.16&$-3.34$&$-3.70$& 0.15$\sim$0.45$^d$\\
 & Li(digl)$_2$C$_{20}$ &4.52&5.31&$-$0.89&~~~0.31&~~~0.99&1.75&$-$2.08&$-$1.01&$<$0.75$^d$\\
\hline
\end{tabular} \\
$^a$ Ref.~\cite{boettiger}, $^b$ Ref.~\cite{graphitexf1,graphitexf2}, $^c$~Ref.~\cite{Kittel}, $^d$ Ref.~\cite{Jache}
\end{table*}

\begin{figure}[!b]
\centering
\includegraphics[clip=true,scale=0.52]{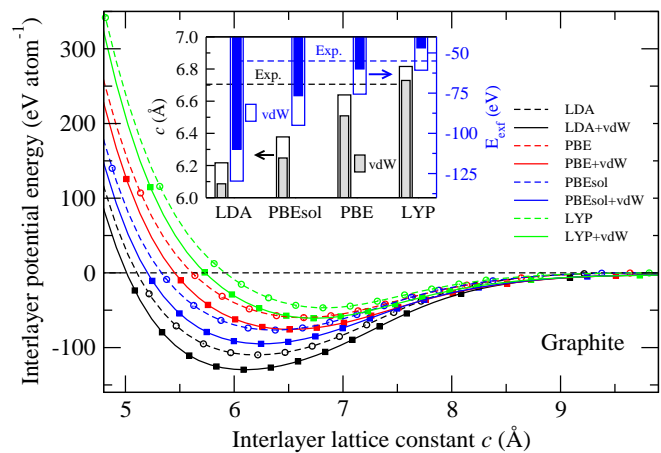}
\caption{The interlayer potential energy per atom in a graphite unit cell as a function of the interlayer lattice constant $c$ obtained from different XC functionals with and without vdW correction. The inset shows the equilibrium lattice constants and the exfoliation energies according to the XC functionals.}
\label{fig_graexf}
\end{figure}

We then tested graphite unit cell involving 4 carbon atoms, of which lattice constants and exfoliation energy per carbon atom were computed with the different XC functionals. Here, the equilibrium interplanar lattice constant $c$ and the exfoliation energy $E_\text{exf}$ were determined by calculating the interlayer potential energies as a function of the length of lattice constant $c$ and then extracting the position where the interlayer potential energy is in minimal. The in-plane lattice constant $a$, governed by the C$-$C covalent ($\sigma$ and $\pi$) bonding, is well reproduced with all XC functionals as listed in Table~\ref{tab_xcene}. However, the interlayer distance $c$ along the direction of vdW interactions depends evidently on the choice of XC functional as listed in Table~\ref{tab_xcene} and shown in Fig.~\ref{fig_graexf}.
All the XC functionals except LYP, with localized DZP$-$PAO basis sets, yielded distinct underestimation of lattice constant $c$ unlike plane-wave basis sets, with which PBE gave the significant overestimation, LDA the underestimation, and PBE+vdW could give proper value~\cite{Tsai,yucj04,yucj06}. Inclusion of vdW correction lowered the lattice constant by $\sim$0.12 \AA, and the exfoliation energy by 15$\sim$20 meV. It was found that LYP+vdW yields the closest interlayer lattice constant $c$ of 6.730 \AA~to the experimental value of 6.705 \AA~\cite{boettiger} and LYP the best exfoliation energy of $-47$ meV to the experimental value of $-35\sim-52$ meV~\cite{graphitexf1,graphitexf2}. Tasaki~\cite{Tasaki} reported the calculated value of $-69$ meV with PBE+vdW using double numerical polarization atomic orbital basis sets, while Dappe {\it et al.}~\cite{graphitexf3} obtained $-61\sim-74$ meV through a combination of the DFT approach and the perturbation theory. Their results are close to those from PBE and PBEsol functionals in this work.
\begin{figure*}[!ht]
\centering
\begin{tabular}{rrr}
\includegraphics[clip=true,scale=0.35]{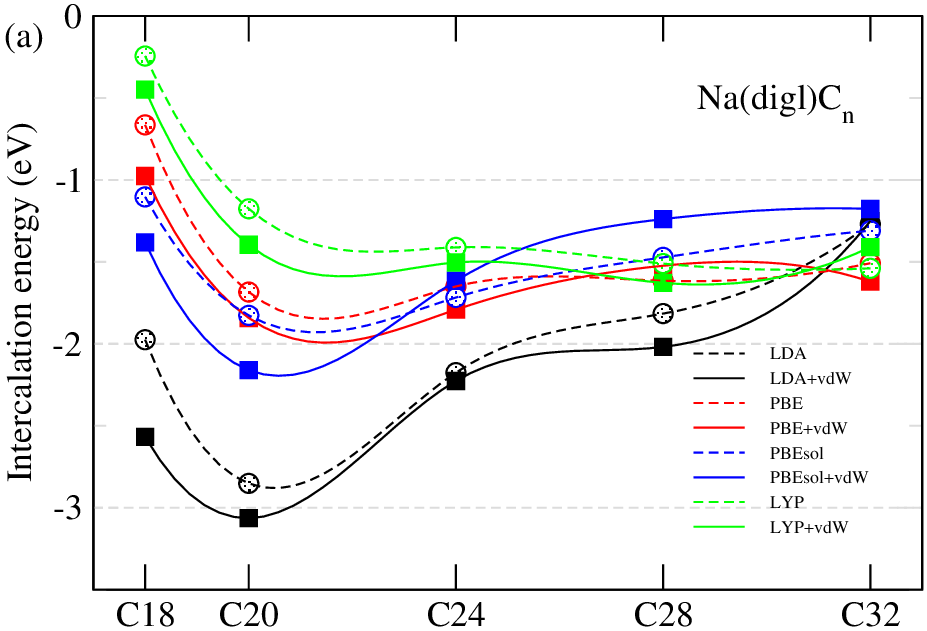} &
\includegraphics[clip=true,scale=0.35]{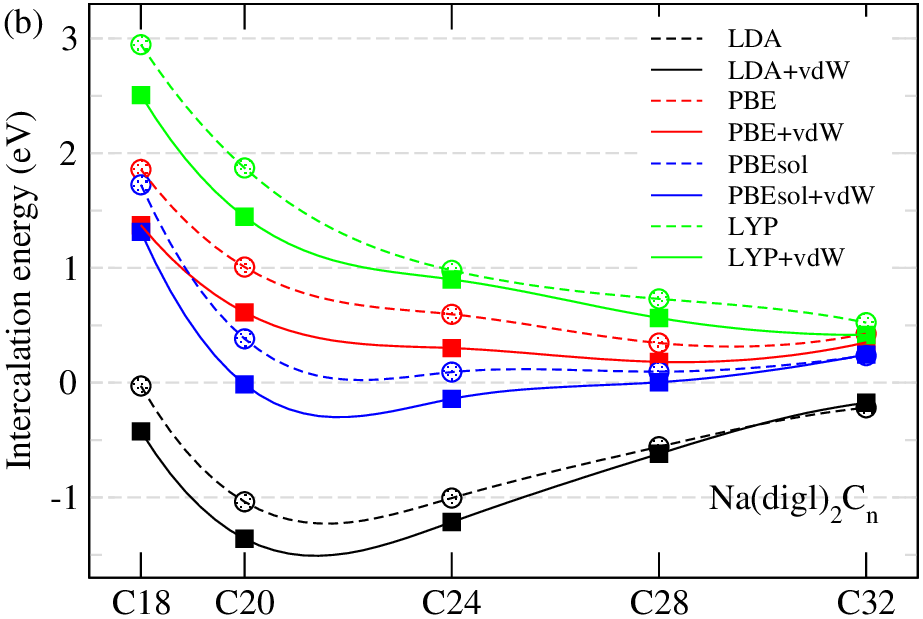} &
\includegraphics[clip=true,scale=0.35]{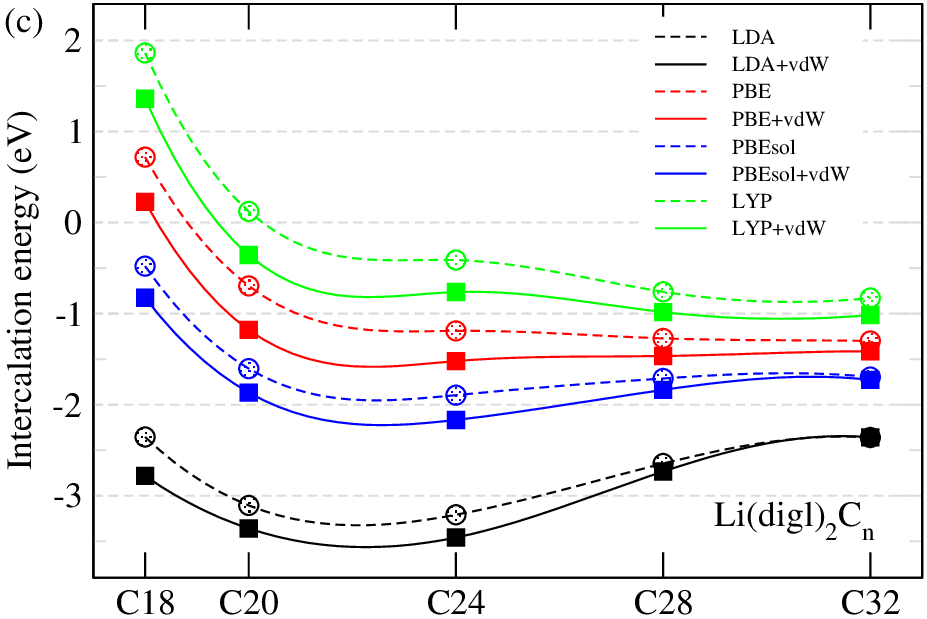} \\ \\
\includegraphics[clip=true,scale=0.35]{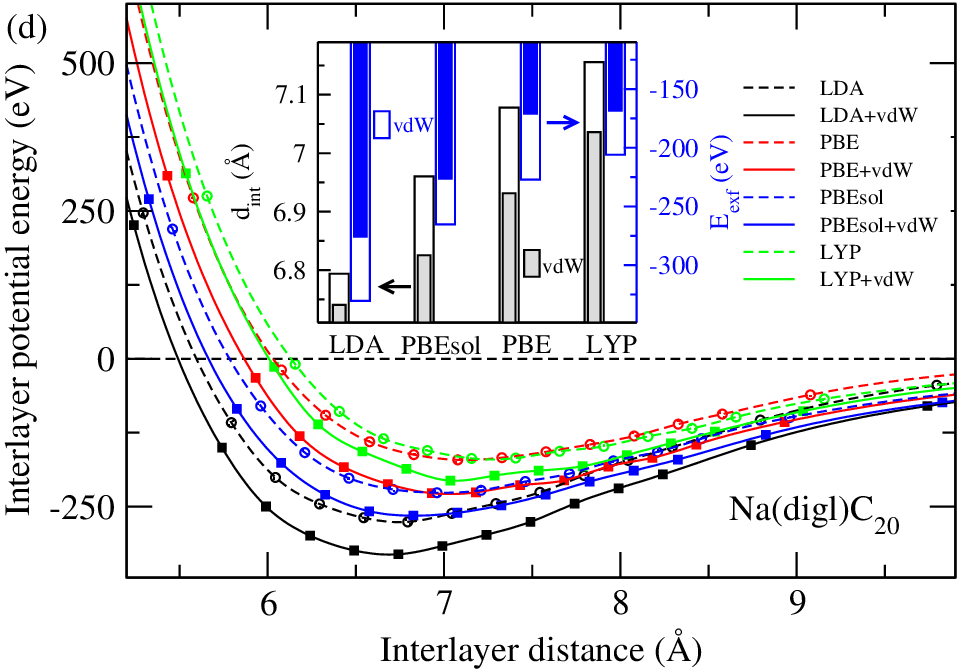} &
\includegraphics[clip=true,scale=0.35]{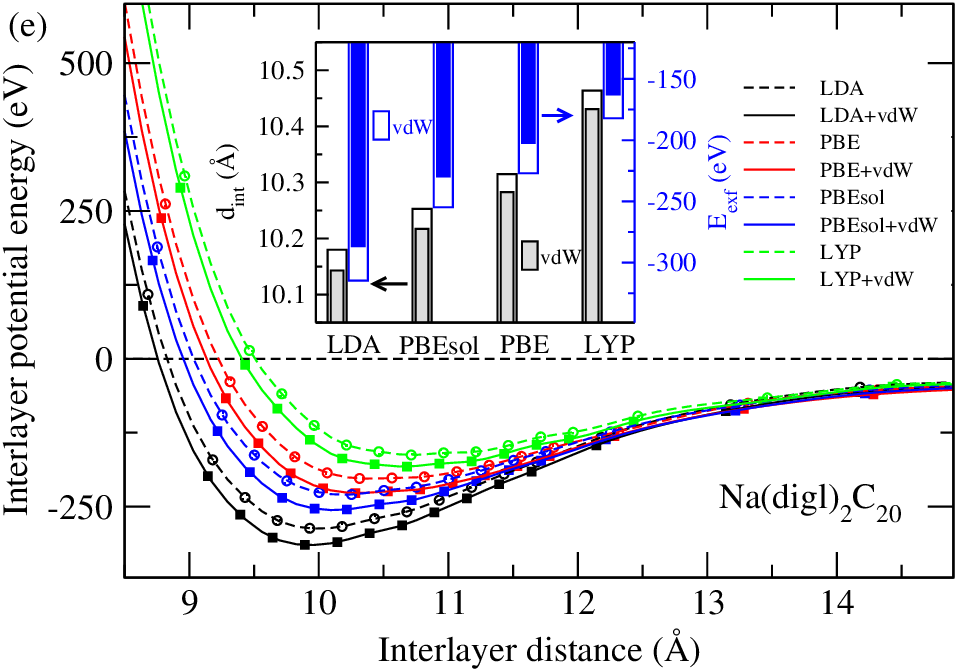} &
\includegraphics[clip=true,scale=0.35]{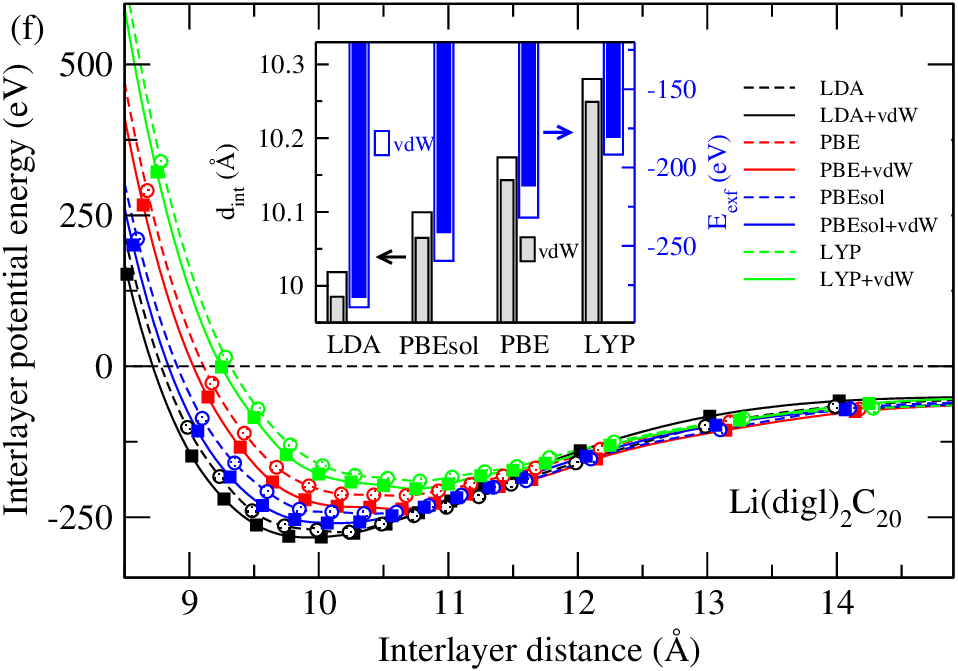} \\ \\
\end{tabular}
\caption{The intercalation energy as a function of carbon atom number in (a) Na(digl)C$_n$, (b) Na(digl)$_2$C$_n$, and (c) Li(digl)$_2$C$_n$, and the interlayer potential energy as a function of interlayer distance in (d) Na(digl)C$_{20}$, (e) Na(digl)$_2$C$_{20}$, and (f) Li(digl)$_2$C$_{20}$, obtained from different XC functionals with and without vdW correction. The insets show the equilibrium interlayer distances and the exfoliation energies.}
\label{fig_EneXC}
\end{figure*}

For further assessment of the XC functionals (and vdW correction) and getting an insight of $t$-GIC formation (and stability), the intercalation and exfoliation energies were calculated. Among different configurations, the lowest energetic configuration was determined by comparing their total energies (see Fig.S2\dag). As shown in Figs.~\ref{fig_EneXC}(a)-(c), all the XC functionals produce the negative intercalation energies for almost Na(digl)C$_n$ and Li(digl)$_2$C$_n$ compounds, while only LDA, LDA+vdW and PBEsol+vdW functionals could give the negative values for Na(digl)$_2$C$_n$ compounds with some $n$ values. It should be emphasized that with PBEsol+vdW functional the lowest intercalation energy in Na(digl)$_2$C$_n$ was found at $n\approx21$ ({\it i.e.}, the most stable compound is Na(digl)$_2$C$_{21}$), corresponding to specific capacity of $\sim$97 mAh g$^{-1}$ close to the experimentally identified capacity of $\sim$100 mAh g$^{-1}$~\cite{Jache}. In addition, when including the vdW correction, the intercalation energies become slightly lower in these compounds. We see from Figs.~\ref{fig_graexf} and~\ref{fig_EneXC}(d)-(f) that the interlayer distance $d_\text{int}$ is calculated to become larger and the magnitude of exfoliation energy smaller going from LDA to PBEsol, PBE and LYP. These indicate that LDA induces more binding than GGA and the inclusion of vdW correction leads to stronger binding in Na/Li $t$-GICs.

The electrode voltage represents one of crucial electrochemical properties to determine the suitability of material as an electrode for NIBs. Assuming that during the charge process Na-GICs transform from the compound of higher carbon atom concentration ({\it i.e.}, lower Na concentration compound) to the compound of lower carbon atom concentration (higher Na concentration), we convert the $t$-GICs from Na(digl)$_x$C$_n$ to Na$_{x_i}$(digl)$_{2x_i}$C$_{32}$ with $x_i=32/n$, treating C$_{32}$ $t$-GIC as the reference compound (starting compound). Then, the electrode potential $V_\text{el}$ with respect to Na/Na$^+$ can be calculated as follows~\cite{Aydinol},
\begin{equation}
\label{eq_Vel}
V_\text{el}=-\frac{E_{\text{Na}_{x_j}(\text{digl})_{2x_j}\text{C}_{32}}-E_{\text{Na}_{x_i}(\text{digl})_{2x_i}\text{C}_{32}}-(x_j-x_i)(2E_\text{digl}+E_{\text{Na}_{\text{bcc}}})}{(x_j-x_i)e}
\end{equation}
In Table~\ref{tab_xcene}, the electrode voltages calculated with $x_i=32/32$, and $x_j=32/24$ for Na $t$-GICs and $x_j=32/20$ for Li $t$-GICs are presented. It was found that PBEsol+vdW give the closest electrode voltage of 0.16 V for Na(digl)$_2$C$_n$ to the experimental values of 0.15 (50th cycle) $\sim$0.45 V (first cycle) at the capacity of 86 mAh g$^{-1}$ corresponding to C$_{24}$ compound~\cite{Jache}.

The most reliable XC functional was determined as a compromise to simultaneously describe the binding properties of graphite and sodium metal, the intercalation energies of Na(digl)$_x$C$_n$ compounds and more importantly, the electrode voltage of Na(digl)$_2$C$_n$ that greatly dominates the electrochemical performance of NIB. As a result, PBEsol+vdW was chosen as the most reliable XC functional since it results in excellent agreement with the electrode voltage of Na(digl)$_2$C$_n$, negative intercalation energy and reasonable binding properties of graphite and Na (Li) metal. This is agreed with that PBEsol can reliably predict material properties of Na-amine $t$-GICs with the plane wave method~\cite{yucj06}.

\subsection{\label{subsec-geo}Variation in structural and energetic properties}
We compared the binding properties of different solvated ions Na(digl), Na(digl)$_2$ and Li(digl)$_2$, which are formed in ether-based electrolyte solution, intercalate into graphite and diffuse in the interlayer space between graphene sheets. From the calculated binding energies as listed in Table~\ref{tab_xcene}, it can be said that the Na atom is less binding to two diglyme molecules in Na(digl)$_2$ ($E_\text{b}=-$0.11 eV) than Na binding to one diglyme molecule in Na(digl) ($-$0.32 eV), and also slightly less than Li binding to two diglyme molecules in Li(digl)$_2$ ($-$0.13 eV). Considering that Na atom is coordinated by three or six oxygen atoms, the average Na$-$O bond length in Na(digl)$_2$ (2.617 \AA) is estimated to be much shorter than that in Na(digl) (2.931 \AA), and also slightly shorter than that in Li(digl)$_2$ (2.697 \AA), as depicted in Fig.~\ref{fig_freemol}. This indicates that stronger binding of Na to diglyme molecule does not induce smaller Na$-$O bond length, implying neither ionic nor covalent bonding. The bonding is actually coordinate bonding.
\begin{figure*}[!th]
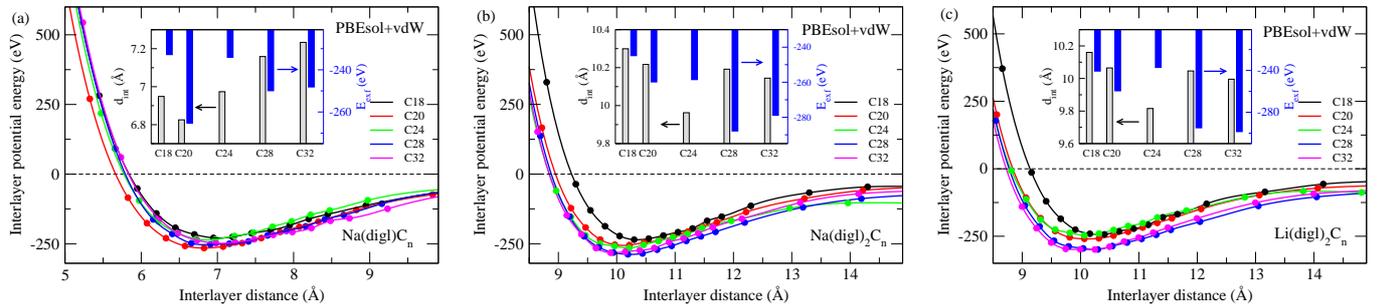

\centering
\begin{tabular}{rrr}
\includegraphics[clip=true,scale=0.35]{fig5a.eps} &
\includegraphics[clip=true,scale=0.35]{fig5b.eps} &
\includegraphics[clip=true,scale=0.35]{fig5c.eps} \\
\end{tabular}
\caption{The interlayer potential energy as a function of interlayer distance in (a) Na(digl)C$_n$, (b) Na(digl)$_2$C$_n$, and (c) Li(digl)$_2$C$_n$ with increasing the number of carbon atoms in graphene sheet, obtained from PBEsol+vdW functional.}
\label{fig_Eexf}
\end{figure*}
\begin{table}[!b]
\caption{Graphene unit cell, specific capacity ($C$), average bond length between Na and O ({\it \={d}}$_\text{Na-O}$), interlayer distance ($d_\text{int}$), relative volume expansion ratio ($r_\text{vol}$), exfoliation energy ($E_\text{exf}$), and intercalation energy ($E_\text{int}$) in $t$-GICs Na(digl)$_2$C$_n$ with $n=$18, 20, 24, 28, and 32.}
\label{tab_pbesolvdw}
\begin{tabular}{l@{\hspace{5pt}}c@{\hspace{5pt}}c@{\hspace{5pt}}c@{\hspace{5pt}}c@{\hspace{5pt}}c}
\hline
     & C$_{18}$ & C$_{20}$ &C$_{24}$ &C$_{28}$ &C$_{32}$ \\
\hline
Unit cell & \small $(3\times3)$ &\small $(\sqrt{12}\times\sqrt{7})$&\small $(4\times3)$ &\small $(\sqrt{12}\times\sqrt{13})$&\small $(4\times4)$ \\
$C$ (mAh g$^{-1}$)   &112.0&101.8&86.1&74.6&65.8\\
{\it \={d}}$_\text{Na-O}$ (\AA) & 2.253&2.315&2.275&2.297&2.277 \\
$d_\text{int}$ (\AA) &10.30&10.22&9.96&10.19&10.15 \\
$r_\text{vol}$ (\%)  &240&236&223&231&229 \\
$E_\text{exf}$ (meV) &$-$245&$-$260&$-$259&$-$288&$-$279\\
$E_\text{int}$ (eV)  &1.31&$-$0.02&$-$0.14&0.00&0.24\\
\hline
\end{tabular}
\end{table}
When the solvated A(digl)$_x$ ion intercalates into graphite, the binding between graphene sheets becomes stronger due to the additional interaction between the intercalants and the graphene sheet, as evidenced by that the exfoliation energies in these $t$-GICs are calculated to be negatively larger than that in graphite ($-$95 meV per carbon atom). Of the three different $t$-GICs at $n=20$, as listed in Table~\ref{tab_xcene}, Na(digl)$_2$C$_{20}$ ($-$255 meV) has the smallest exfoliation energy in magnitude, while Na(digl)C$_{20}$ ($-$265 meV) the largest, and Li(digl)$_2$C$_{20}$ ($-$259 meV) in-between, indicating that the binding between graphene sheets in Na(digl)$_2$C$_n$ is weaker than in Na(digl)C$_n$ and also Li(digl)$_2$C$_n$. However, we should note that the differences between binding energies are not very significant, and moreover, due to an uncertainty of selecting $d_\text{i}=\infty$ ($\sim$30 \AA~in this work) in Eq.~\ref{eq_Eexf} and some degrees of errors in the calculations, the differences are almost negligible. On the other hand, as can be seen in Fig.~\ref{fig_Eexf}, when changing the number of carbon atoms of graphene sheet in these $t$-GICs, the exfoliation energy changes wavily as the interlayer distance dose so. In the cases of intercalants with two diglymes, the interlayer distance is minimal at $n=24$, while the exfoliation energy is also minimal in magnitude, indicating an unrelation between binding strength and interlayer distance. 

\begin{figure}[!b]
\centering
\includegraphics[clip=true,scale=0.52]{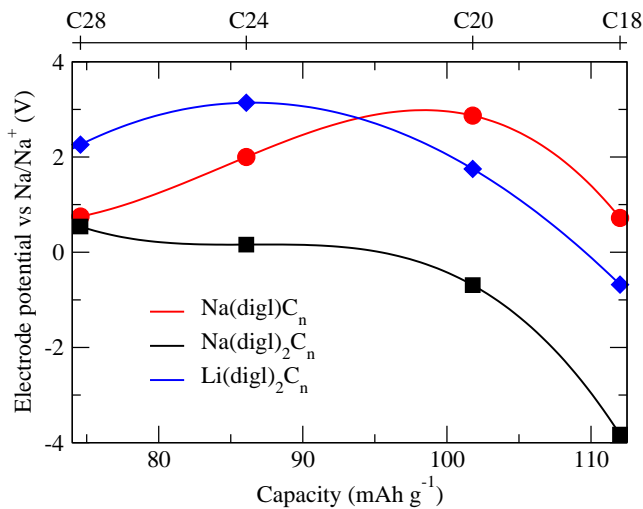}
\caption{The electrode potential versus Na/Na$^+$ as a function of capacity in Na(digl)C$_n$, Na(digl)$_2$C$_n$ and Li(digl)$_2$C$_n$, evaluated as average potential treating C$_{32}$ compound as reference.}
\label{fig_Vel}
\end{figure}
For further insight into the structural variation upon the intercalation of the solvated Na(digl)$_2$ ion into the graphite, we considered the average Na$-$O bond length and the relative volume expansion ratio, $r_\text{vol}=(V-V_0)/V_0\times100$ (\%), together with the interlayer distance. As listed in Table~\ref{tab_pbesolvdw} and shown in Fig.~\ref{fig_model}, the average Na$-$O bond lengths (2.253$\sim$2.315 \AA) in Na(digl)$_2$ intercalants that are constrained between graphene sheets are remarkably contracted compared to that (2.617 \AA) in free solvated complexes, indicating an enhancement of Na$-$diglyme binding upon its intercalation into graphite (see Fig.S3 for Na(digl)C$_n$\dag). This might be due to somewhat complicated interactions between Na atom, diglyme molecules and graphene sheets. It is worthy noting that such enhancement of Na binding to the solvent can prevent decomposition of solvated ion, resulting in formation of a negligible solid-electrolyte interphase (SEI) film on the graphite surface, enabling Na$^+$-solvent migration into the graphite lattice~\cite{Kim15}. On the other hand, the interlayer distances $d_\text{int}$ are almost over 10 \AA, much larger than that of graphite (3.353 \AA). Compared with experiments, Jache and Adelhelm reported a 10\% expansion for Li intercalation and a 15\% expansion for sodium~\cite{Jache}, while Kim {\it et al.}~\cite{Kim152} reported 11.62 \AA~for Na$_2$(digl)$_2$ cointercalated compound. Therefore, our calculated interlayer distances are reasonable compared with the latter experiment, though the number of intercalated sodium atoms is different. At this point, we note that the interlayer distances are more or less constant with the maximum relative change of 3\% as increasing the number of carbon atoms of graphene sheet. This implies that, once $t$-GIC is formed in the first cycle, there would be no serious change of interlayer distance and volume as shown in Table~\ref{tab_pbesolvdw} (Table S1\dag~for Na(digl)C$_n$ and Li(digl)$_2$C$_n$), during further cycles.
\begin{figure*}[!t]
\centering
\begin{tabular}{cc}
\includegraphics[clip=true,scale=0.23]{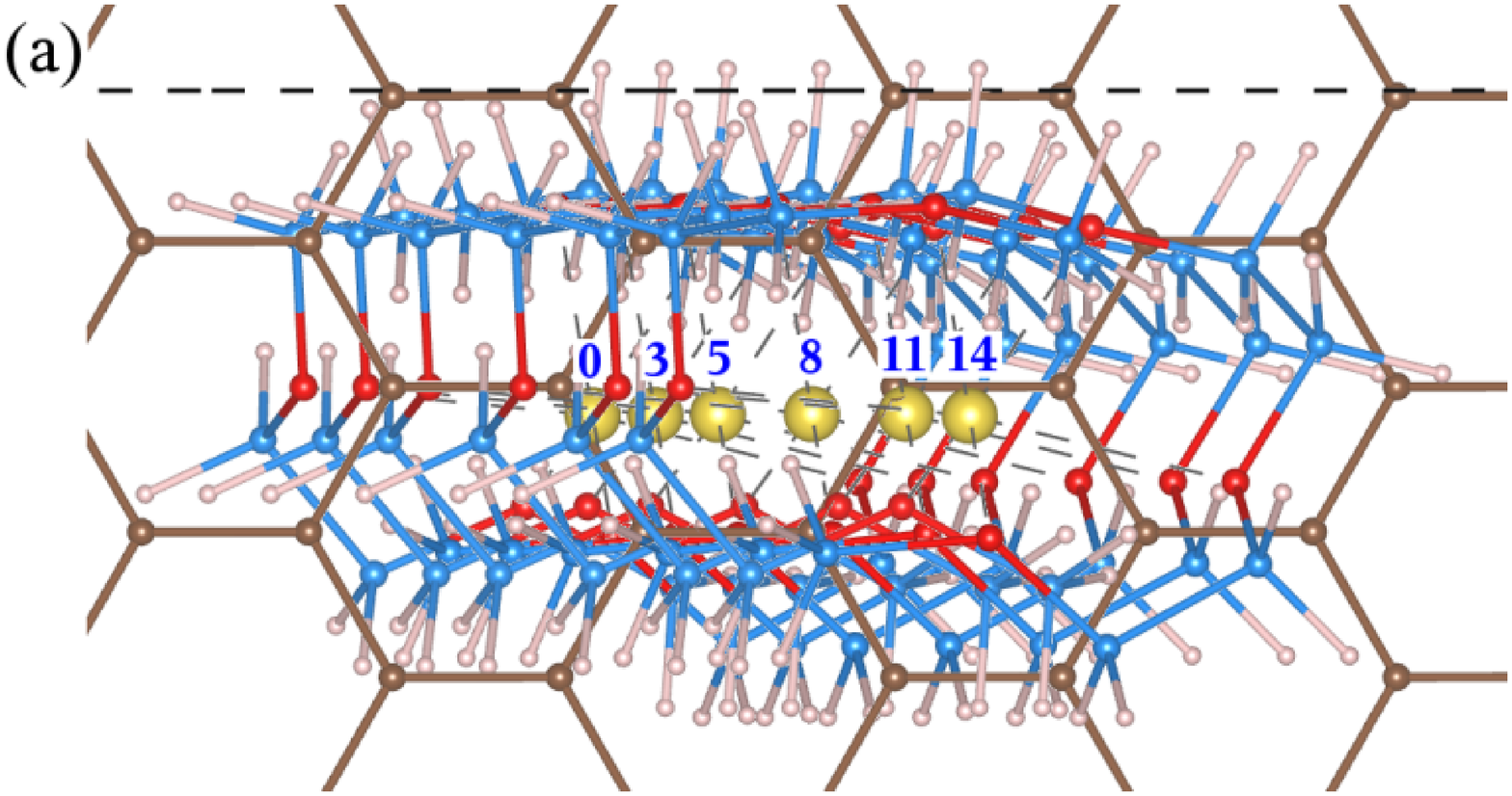} \hspace*{30pt} & \hspace*{20pt}
\includegraphics[clip=true,scale=0.23]{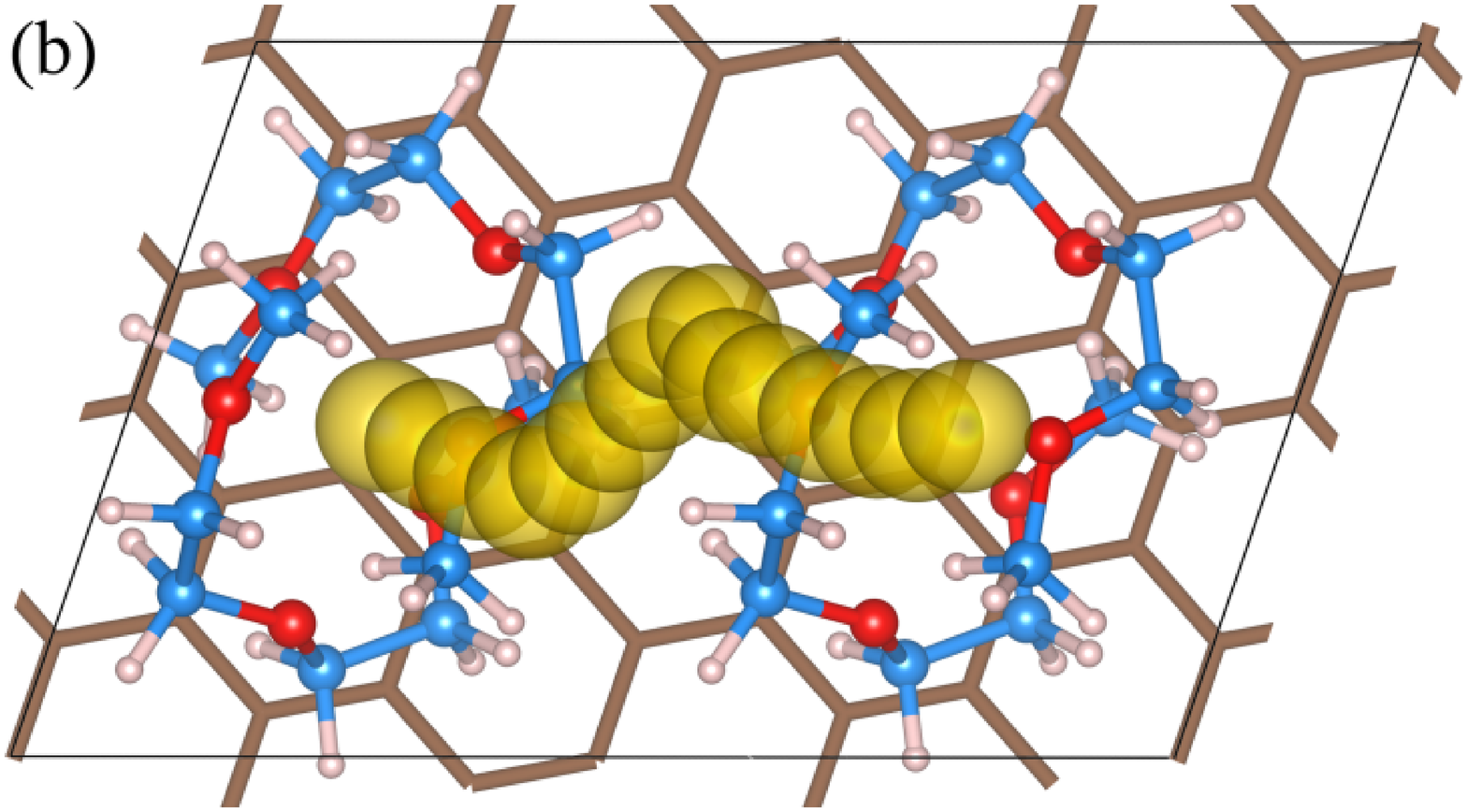} \\
\includegraphics[clip=true,scale=0.5]{fig7c.eps} \hspace*{40pt} &
\includegraphics[clip=true,scale=0.5]{fig7d.eps} \\
\end{tabular}
\caption{Migration pathways for (a) Na(digl)$_2$ solvated ion diffusion and (b) Na ion extraction, and the corresponding activation energies for (c) Na(digl)$_2$ solvated ion and (d) Na ion extraction in Na(digl)$_2$C$_{20}$ compound, respectively.}
\label{fig_mig}
\end{figure*}

As listed in Table~\ref{tab_pbesolvdw} and discussed above, the intercalation energies for Na(digl)$_2$C$_n$ compounds are negative only for $n=20\sim28$, as far as we rely on the PBEsol+vdW functional calculations. This is agreed well with the experiment that Na(digl)$_2$C$_{20}$ is preferably formed~\cite{Jache}. When one diglyme molecule instead of two molecules or Li atom instead of Na atom intercalates into graphite to form Na(digl)C$_n$ or Li(digl)$_2$C$_n$, the intercalation energies become negative with larger magnitude for almost $n$ values by all the XC functionals. This indicates that it is easier to form Na-diglyme intercalated graphite with one diglyme molecule than with two diglyme molecules, and to form Li-GICs than Na-GICs. Nevertheless, the change tendency of intercalation energies in Li(digl)$_2$C$_n$ as increasing the number of carbon atoms is similar to Na(digl)$_2$C$_n$.  To check the suitability of these compounds for electrode, we computed the electrode potential as a function of capacity by applying Eq.~\ref{eq_Vel}, as shown in Fig.~\ref{fig_Vel}. It is found that Na(digl)$_2$C$_n$ has the lowest electrode potential below 0.5 V for the range of capacities considered in this work, indicating it's the most suitable anode material among three kinds of $t$-GICs.

\subsection{\label{subsec-elechem}Ionic diffusion and electronic conduction}
The fundamentals of sodium ion transport such as ionic diffusion pathways and their activation barriers are an important aspect for charge-discharge kinetics, and in some cases this is an critical issue for the cycle stability. In fact, the activation energy for sodium ion diffusion during cycling is a determinant factor how fast charge process takes place, and diffusion pathways should be short to enable fast migration of Na ions during battery operation. In this sense, we inquire into the diffusion process of the solvation ions with a calculation of activation energies by applying the nudged elastic band (NEB) method~\cite{NEB} as implemented in Python script Pastafarian\footnote{This code was originally developed by J. M. Knaup, and we modified the code to debug some minor errors and allow parallel running with a permission.} in conjunction with SIESTA code. We typically choose A(digl)$_x$C$_{20}$ compounds, and utilized doubled unit cells in the $b$-direction (doubled unit cell) to allow migrations of only alkali ions and in both $a$ and $b$-directions (quadrupled unit cell) for A(digl)$_x$ ions from one position to another identical position. The number of NEB images to discretize the path are 15$\sim$17, and atomic relaxations were allowed during the NEB search.

In Fig.~\ref{fig_mig}(a) and (c), we show the calculated activation energies for migrations of the Na(digl)$_2$ solvated ions along the shortest pathways, with a depiction of geometries of some transition states. The activation energy of this migration in Na(digl)$_2$C$_{20}$ compound was calculated to be about 0.40 eV, which is slightly higher than that in Na(digl)C$_{20}$ (0.28 eV) and that in Li(digl)$_2$C$_{20}$ (0.25 eV) (see Fig.S4 for Na(digl)C$_{20}$ and Li(digl)$_2$C$_{20}$\dag). The latter comparison indicates that Li solvated ions diffuse more smoothly than Na solvated ion, which is contrast to the previous {\it ab initio} calculations for different $t$-GICs~\cite{yucj06} as well as for alkali atom $b$-GICs~\cite{Nobuhara}. It can be said in safe that the $t$-GICs investigated in this work can offer reasonably fast charge-discharge rate since the activation energies are comparable with Li diffusion barrier in graphite (0.40 eV)~\cite{Nobuhara} and Na diffusion barriers in layered anode materials like MoS$_2$ (0.28 eV)~\cite{Mortazavi} and cathode materials like NaCoO$_2$ (0.3 eV)~\cite{Ong}. We also looked over migration of only sodium ion, which might be helpful for drawing an inference whether the formation of SEI film is easy or not. As shown in Fig.~\ref{fig_mig}(b) and (d), the activation energy for this migration was calculated to be much high over 3 eV, indicating that the SEI film formation is not easy.
\begin{figure}[!t]
\centering
\includegraphics[clip=true,scale=0.6]{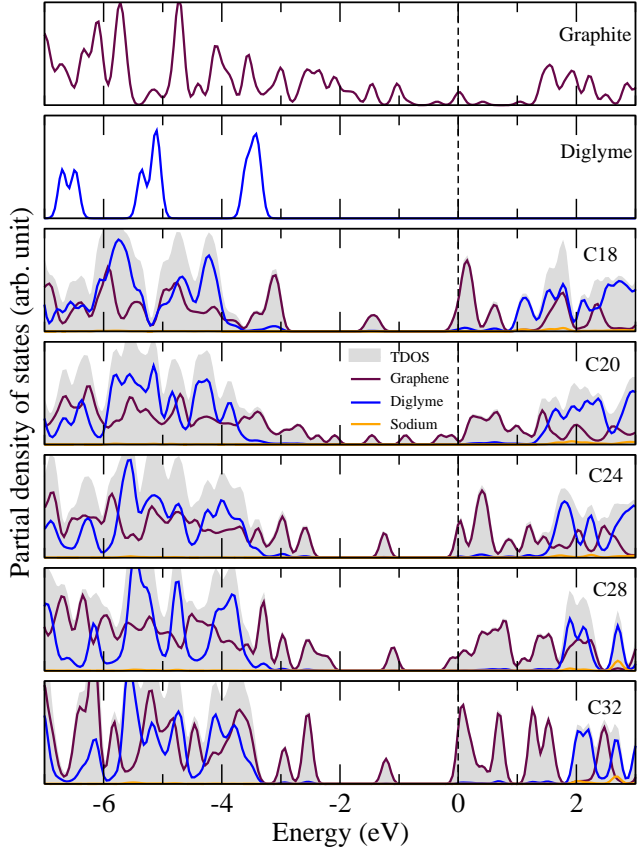}
\caption{Partial density of states in free diglyme molecule and Na(digl)$_2$C$_n$ compounds with $n=$18, 20, 24, 28 and 32, calculated by using PBEsol+vdW functional. Fermi energies are set to be zero, indicated by dashed vertical lines.}
\label{fig_dos}
\end{figure}

We then turn to a question how electrons are conducted inside materials during charge-discharge process. This is an important issue for understanding the battery operation since the electrons are transmitted to/from the collector the moment the solvated sodium ions de/intercalates out from/into the graphite. To qualitatively estimate the electron conductance of these materials, we analyse the density of states (DOS) of the relevant compounds. In Fig.~\ref{fig_dos}, we show the total DOS (TDOS) and atomic resolved partial DOS (PDOS) of major spin component (those of minor spin are almost identical to those of major spin) of graphite, isolated diglyme molecule, and Na(digl)$_2$C$_n$ compounds. It was well known that graphite has excellent in-plane electric conductance (no inter-plane conductance), whose charge carrier is just $p_z$ valence electron of carbon~\cite{Dresselhaus,yucj06}. When the solvated sodium ion intercalates into graphite, the in-plane electronic conductance is retained well, even better than graphite, as the density of electronic states of carbon around Fermi level in $t$-GICs is relatively higher than in graphite. In Fig.~\ref{fig_dos}, we see the unoccupied state of sodium around 3 eV over Fermi level, indicating that sodium atom donates its $s$ electron to become a cation (electron donor) while carbon atoms in graphene layer accept it to occupy anti-bonding $\pi^*$  ($p_z$) orbital (electron acceptor). We also observe significant orbital overlaps between graphene layer and diglyme below and above Fermi level, indicating hybridization between $\sigma$ bonding orbitals of carbon atoms in graphene layer and molecular orbitals of diglyme molecule, resulting strong interaction between the electrons. Similar characteristics is observed in Na(digl)C$_n$ and Li(digl)$_2$C$_n$ compounds (Fig.S5(a) and (b)\dag).

\subsection{\label{subsec-electro}Formation mechanism of compounds}
It is generally accepted that charge transferring is occurred in the formation of GIC from its components. Therefore, consideration of charge transferring is indispensable to make it clear the nature of chemical bonding and thus the formation mechanism. As discussed above, this could be done by analysing the electronic structure such as DOS. To get more intuitive insight into the charge transfer, we investigate a distribution of the charge density difference in Na-solvent cointercalated graphite compounds as calculated by using the following equation,
\begin{equation}
\label{eq_rho}
\Delta n(\textbf{\textit{r}})=n_{\text{Na(digl)}_x\text{C}_n}(\textbf{\textit{r}})-[n_{\text{Na}}(\textbf{\textit{r}})+n_{\text{(digl)}_x}(\textbf{\textit{r}})+n_{\text{C}_n}(\textbf{\textit{r}})]
\end{equation}

\begin{figure}[!t]
\centering
\includegraphics[clip=true,scale=0.39]{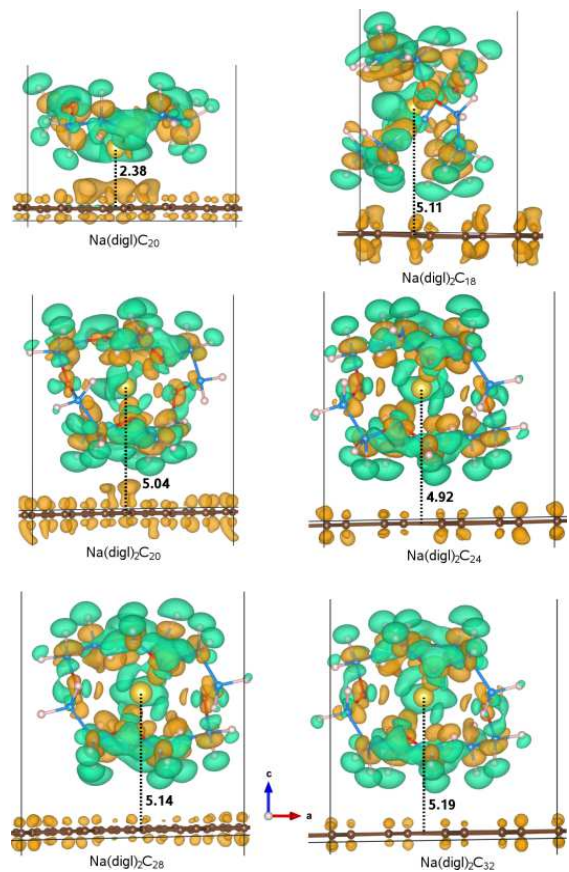}
\caption{Isosurface plot of charge density difference between Na(digl)$_x$C$_n$ and the sum of graphene layer, sodium atom and diglyme molecule, evaluated at $\pm$0.0016 $|e|$~\AA$^{-3}$. Brown-coloured isosurface represents positive value (electron gain), while green-coloured negative value (electron loss).}
\label{fig_dens}
\end{figure}

Fig.~\ref{fig_dens} shows the electronic density accumulation (brown colour) and depletion (green colour) in Na(digl)C$_{20}$ and Na(digl)$_2$C$_n$ with $n=$18, 20, 24, 28, 32 (isosurface of $\pm$0.0016 $|e|$~\AA$^{-3}$). In overall, the Na atom has lost its valence electronic density, and the solvent molecules have parts of electron loss and gain together, while the graphene layer has gained the electrons. Observing the distribution of charge density difference does not tell whether the solvent molecule in the aggregate has lost electrons or not. In the case of Na(digl)$_x$C$_n$, some part of electrons has transferred from Na atom to the bonding carbon rings, and this induces the formation of ionic bond between Na ions and carbons, with electrons donated from C$-$C bonds in the graphene layers (Fig.S6(a) and (b)\dag). It is found that as increasing the number of carbon atoms in graphene layer the amount of accepted electrons becomes smaller. It should be noted that in the case of two diglyme molecules intercalation the Na atom could not interact directly with graphene layer due to a screening by atoms of solvent molecule.

We performed Mulliken population analysis of these compounds to quantitatively estimate the amount of transferred charge, although it is said that such analysis could not yield accurate values for atomic or molecular charges. It turns out that the Na atom in Na(digl)$_x$C$_n$ compounds donates a relatively large amount of electrons (0.142$\sim$0.191 $|e|$), but contrastingly the Li atom in Li(digl)$_2$C$_n$ receives electrons (0.056$\sim$0.074 $|e|$) due to probably weaker electropositivity of Li than Na atom (Table S2\dag). In the entirety, diglyme molecules just release electrons, of which the amount increases in general as increasing the number of carbon atoms in graphene layer, indicating that diglyme molecule is an electron doner. The amount of released electrons from diglyme in Na(digl)$_2$C$_n$ (0.138$\sim$0.233) is larger than in Na(digl)C$_n$ (0.058$\sim$0.122 $|e|$), but smaller than in Li(digl)$_2$C$_n$ (0.342$\sim$0.421 $|e|$). Inside the diglyme molecules, carbon atom releases electrons of $\sim$0.60 $|e|$ in average, hydrogen atom accepts majority of electrons of $\sim$0.25 $|e|$, and oxygen atom accepts minority of electrons of $\sim$0.04 $|e|$, indicating the formation of covalent bond between C and H, and ionic bond between C and O atoms inside the molecule. On the other hand, the amount of accepted electrons per carbon atom of graphene layer decreases in general as increasing the number of carbon atoms, and that in Na(digl)$_2$C$_n$ (0.017$\sim$0.012 $|e|$) is larger than in Na(digl)C$_n$ (0.009$\sim$0.010 $|e|$) but similar to Li(digl)$_2$C$_n$ (0.017$\sim$0.011 $|e|$).

Finally we try to describe the formation mechanism of these $t$-GICs. When forming graphite from carbon atoms, the four valence electrons of carbon make three $sp^2$ hybrid orbitals and one $p_z$ orbital. The three $sp^2$ electrons form strong $\sigma$ bonding in the carbon hexagon on the plane of graphene sheet, while the $p_z$ electron forms weak delocalized $\pi$ bonding with those from other carbon atoms in vertical direction to the hexagon plane. Meanwhile, the graphene layers are weakly binding through the vdW interactions, and thus atoms, molecules or ions can be easily intercalated between the graphene sheets. Depending on the electronegativity of the intercalant, the graphene layer can release the $\pi$ electrons to the intercalant or accept electrons from the intercalant, forming ionic bonding through $\pi$-conjugate interaction between graphene layer and intercalant (covalent bonding can also be formed). When only Na atom intercalates into graphite, the electrons move from the Na atom (strong electropositive element) to the graphene layer but the amount of transferred electrons is relatively so small that the $\pi$ interaction is not enough strong to form a stage-I $b$-GIC. When the diglyme molecules intercalate with Na atom into graphite, the diglyme molecule as well as Na atom donates electron to the graphene layer as discussed above, and therefore the $\pi$-conjugate interaction becomes enough strong to form a stable stage-I GIC. In the case of Li atom, the amount of released electrons from diglyme is so much that the Li atom as well as graphene layer should receive electrons in contrast to the Na case. On the other hand, the graphene layer affects the binding of Na atom to diglyme molecule, such that the Na atom could not easily escape but diffuse in the aggregate fast into the graphite inside. In this context, the structure of cointercalants surrounded by two diglyme molecules could be more favourable than one diglyme molecule. Consequently, the interactions between diglyme molecule and graphene layer play a major role in the formation and operation of these $t$-GICs, while the interaction between sodium atom and graphene layer is rather indirect.

\section{\label{sec:con}Conclusions}
We investigated the cointercalation of sodium and diglyme molecule into graphite aiming to clarify the atomistic structure, energetics and electrochemical properties for Na-ion battery application with {\it ab initio} DFT calculations. The ternary graphite intercalation compounds, Na(digl)$_2$C$_n$ with $n=$18, 20, 24, 28, and 32, were considered with the graphene unit cells of $(3\times3)$, $(\sqrt{12}\times\sqrt{7})$, $(4\times3)$, $(\sqrt{12}\times\sqrt{13})$, and $(4\times4)$. In addition, Na(digl)C$_n$ and Li(digl)$_2$C$_n$ compounds were also considered for reference. We first determined the reliable exchange-correlation functional in connection with vdW inclusion that can reproduce the experimental results for structural and binding properties of graphite and Na/Li metal and for electrode potential of NIB, and that can give the negative intercalation energy of $t$-GICs, as PBEsol+vdW functional. With this functional, the variations in structural properties such as interlayer distance, average Na$-$O bond length and relative volume change, and energetic properties such as exfoliation, intercalation energies and electrode potentials, have been studied systematically, finding out that Na(digl)$_2$C$_n$ has the lowest intercalation energy at $n\approx21$ corresponding to the capacity of $\sim100$ mAh g$^{-1}$. We found that the solvated ion can diffuse relatively fast with the migration barrier of $\sim$0.40 eV and the electron conductance can be enhanced after cointercalation when compared with pristine graphite. The calculation results clarify the mechanism of the formation of these $t$-GICS and reveal new prospects for tailoring innovative anode materials of NIBs based on graphitic materials.

\section*{\label{ack}Acknowledgments}
This work was supported partially from the State Committee of Science and Technology, Democratic People's Republic of Korea, under the fundamental research project ``Design of Innovative Functional Materials for Energy and Environmental Application'' (No. 2016-20). The calculations in this work were carried out on the HP Blade System C7000 (HP BL460c) that is owned and managed by Faculty of Materials Science, \\Kim Il Sung University.

\section*{Appendix A. Supplementary data}
Supplementary data related to this article can be found at URL.

\section*{\label{note}Notes}
The authors declare no competing financial interest.

\bibliographystyle{elsarticle-num-names}
\bibliography{Reference}

\end{document}


\title{Supplementary information -- {\it Ab initio} study of sodium cointercalation with diglyme molecule into graphite}

\author{Chol-Jun Yu\footnote{Corresponding author: Chol-Jun Yu, ryongnam14@yahoo.com}, Song-Bok Ri, Song-Hyok Choe, Gum-Chol Ri, Yun-Hyok Kye, and Sung-Chol Kim \\
\small \it Department of Computational Materials Design, Faculty of Materials Science, Kim Il Sung University \\
\small \it Ryongnam-Dong, Taesong District, Pyongyang, Democratic People's Republic of Korea}
\date{}
\maketitle

\begin{figure}[!th]
\begin{center}
\includegraphics[clip=true,scale=0.4]{figs1a.eps}
\includegraphics[clip=true,scale=0.4]{figs1b.eps}
\end{center}
\caption{\label{fig_conv}DFT total energies of Na(digl)$_2$C$_{20}$ unit cells versus (a) MeshCutoff energy while increasing KgridCutoff parameters systematically, and (b) KgridCutoff length while increasing MeshCutoff energies. The total energy converges to 5 meV per cell with 300 Ry MeshCutoff and 10 \AA~KgridCutoff.}
\end{figure}

\begin{figure*}[!th]
\begin{center}
\begin{tabular}{rr}
\includegraphics[clip=true,scale=0.3]{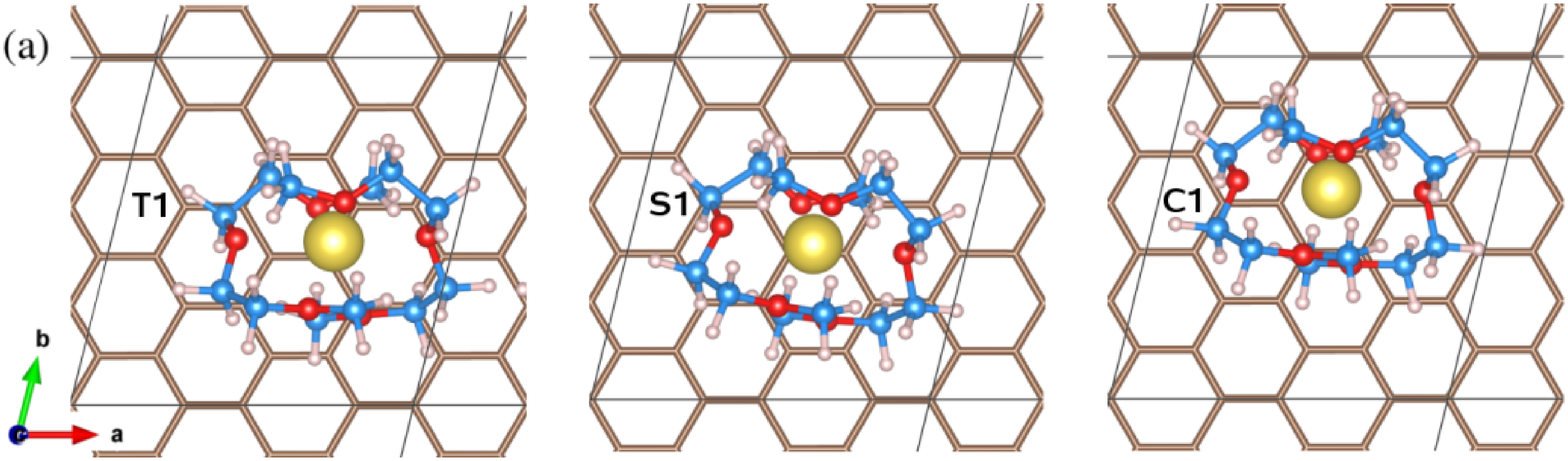} & \includegraphics[clip=true,scale=0.42]{figs2b.eps} \\ 
\includegraphics[clip=true,scale=0.3]{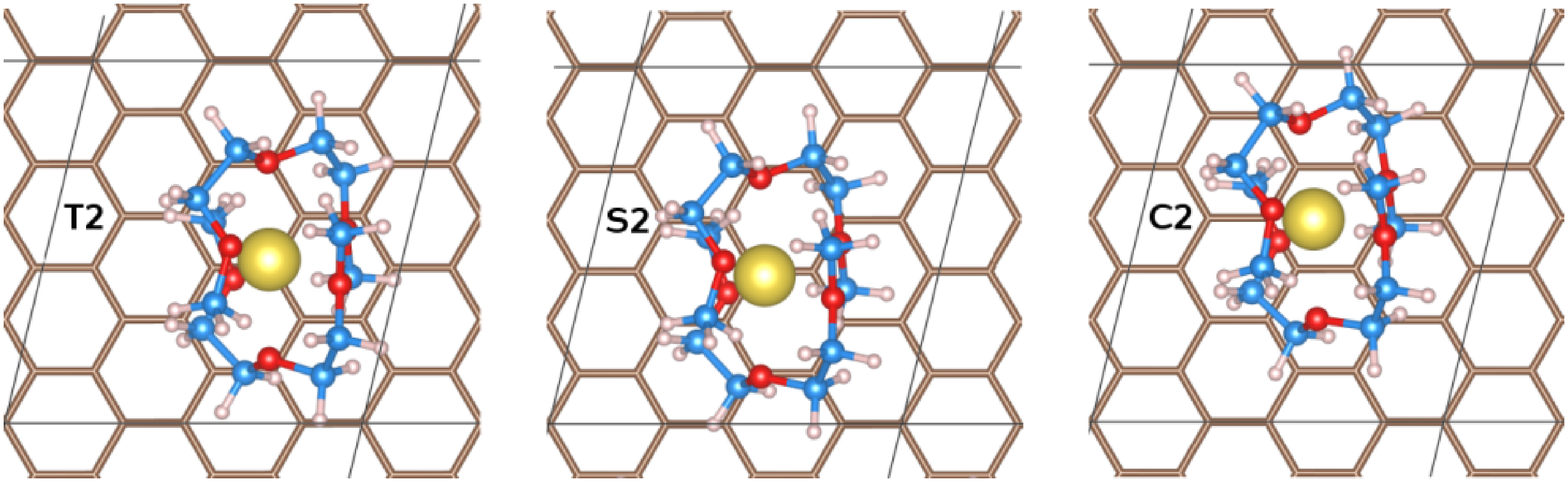} & \includegraphics[clip=true,scale=0.42]{figs2c.eps} \\ 
\end{tabular}
\end{center}
\caption{\label{fig_conf}(a) Top view of different configurations for Na(digl)$_2$C$_{20}$, (b) their total energy differences, and (c) interlayer distances. The sodium positions above graphene sheet are distinguished into triangle point (T), side of hexagon (S) and centre of hexagon (C), and the orientation of face formed by crossing two diglyme molecules. S1 configuration has the lowest energy and the shortest interlayer distance.}
\end{figure*}
\begin{figure*}[!th]
\begin{center}
\includegraphics[clip=true,scale=0.47]{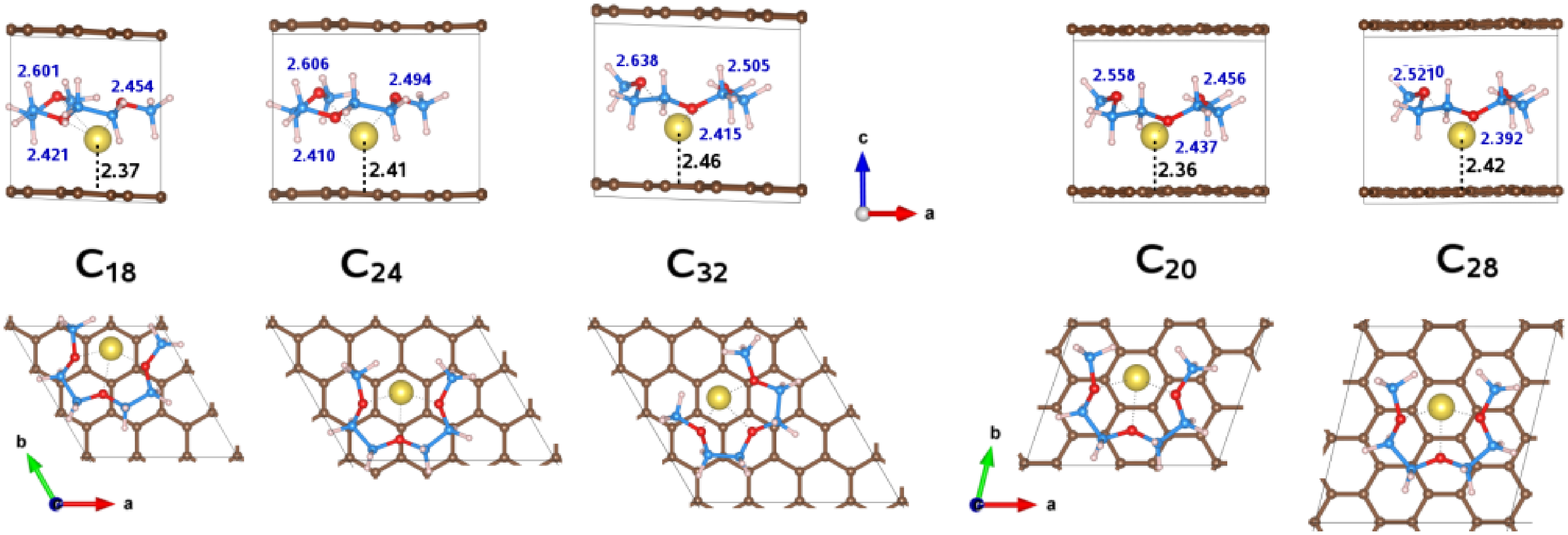}
\end{center}
\caption{\label{fig_model1}Side and top views of unit cells of Na(digl)C$_n$ $t$-GICs, with $n=$18, 24, 32, 20, and 28, optimized by PBEsol+vdW method. Three Na$-$O bond lengths and the length between Na and graphene sheet in \AA~unit are indicated. Unlike Na(digl)$_2$C$_n$, the position of sodium is on the centre of carbon hexagon in graphene sheet.}
\end{figure*}

\begin{table}[!th]
\begin{center}
\caption{Interlayer distance ($d_\text{int}$), average Na/Li$-$O bond length ({\it \={d}}$_\text{Na-O}$), relative volume expansion ratio ($r_\text{vol}$), exfoliation energy ($E_\text{exf}$), and intercalation energy ($E_\text{int}$) of Na(digl)C$_n$ and Li(digl)$_2$C$_n$ ($n=18$, 20, 24, 28, 32).}
\label{tab_main}
\begin{tabular}{cccccc}
\hline
$n$ & $d_\text{int}$ (\AA) & {\it \={d}}$_\text{Na/Li-O}$ (\AA) & $r_\text{vol}$ (\%) & $E_\text{exf}$ (eV) & $E_\text{int}$ (meV) \\
\hline
 & \multicolumn{5}{c}{Na(digl)C$_n$} \\
18 & 6.95 & 2.492 & 127 & $-$233 & $-$1.38 \\
20 & 6.83 & 2.484 & 120 & $-$265 & $-$2.16 \\
24 & 6.97 & 2.503 & 125 & $-$234 & $-$1.61 \\
28 & 7.16 & 2.511 & 130 & $-$250 & $-$1.24 \\
32 & 7.23 & 2.519 & 132 & $-$248 & $-$1.18 \\
\hline
 & \multicolumn{5}{c}{Li(digl)$_2$C$_n$} \\
18 & 10.16 & 2.082 & 234 & $-$243 & $-$0.83 \\
20 & 10.06 & 2.156 & 230 & $-$260 & $-$1.87 \\
24 & ~9.82 & 2.113 & 216 & $-$247 & $-$2.17 \\
28 & 10.05 & 2.128 & 225 & $-$299 & $-$1.84 \\
32 & 10.00 & 2.115 & 223 & $-$299 & $-$1.73 \\
\hline
\end{tabular}
\end{center}
\end{table}

\begin{figure*}[!t]
\begin{center}
\begin{tabular}{cc}
\includegraphics[clip=true,scale=0.5]{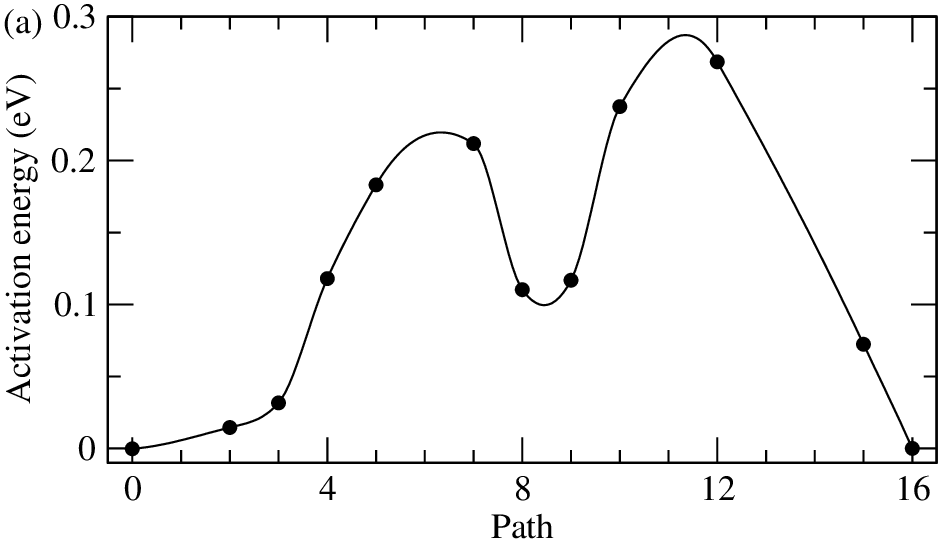} &
\includegraphics[clip=true,scale=0.5]{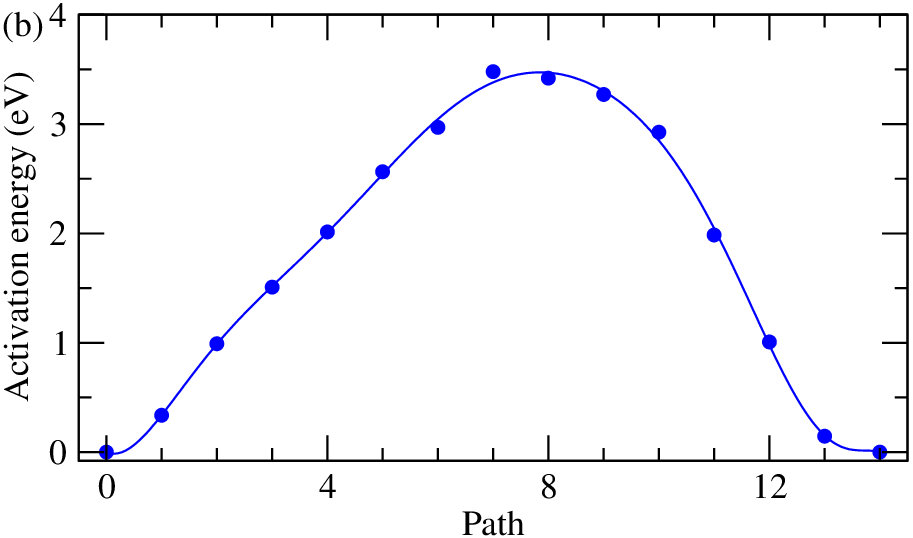} \\
\includegraphics[clip=true,scale=0.5]{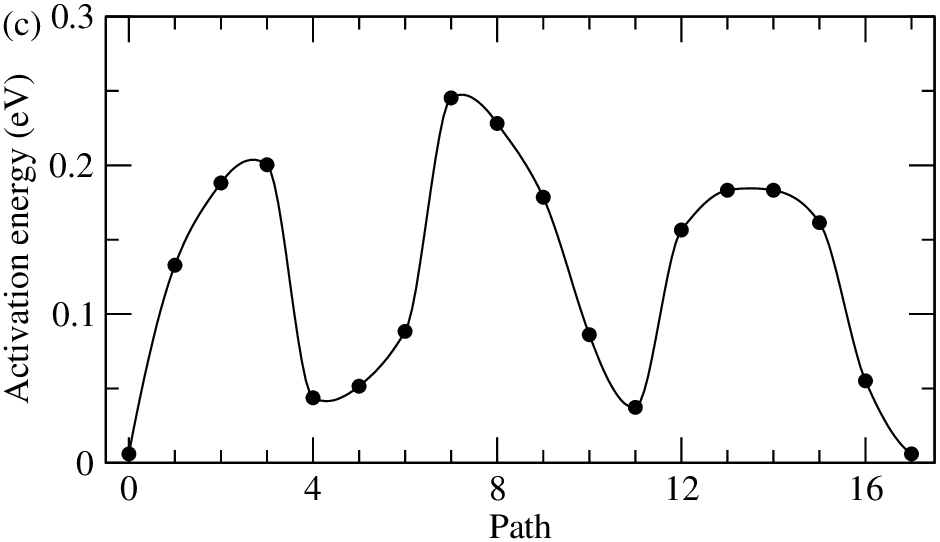} &
\includegraphics[clip=true,scale=0.5]{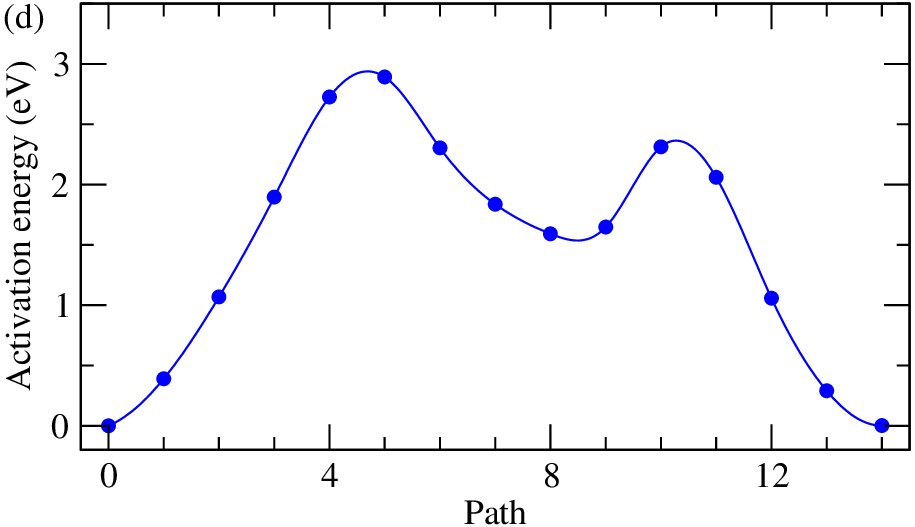} \\
\end{tabular}
\end{center}
\caption{\label{fig_mig}Energy profiles for migrations of (a) Na(digl) and (c) Li(digl)$_2$ solvated ions, and (b) only Na and (d) Li ions, keeping the position of solvent molecules fixed.}
\end{figure*}

\begin{figure*}[!th]
\centering
\includegraphics[clip=true,scale=0.6]{figs5a.eps} \hspace{5pt}
\includegraphics[clip=true,scale=0.6]{figs5b.eps}
\caption{Partial density of states in (a) Na(digl)C$_n$ and (b) Li(digl)$_2$C$_n$ GICs with $n=$18, 20, 24, 28 and 32, calculated by using PBEsol+vdW functional.}
\label{fig_dos}
\end{figure*}
\begin{figure*}[!t]
\begin{center}
\includegraphics[clip=true,scale=0.47]{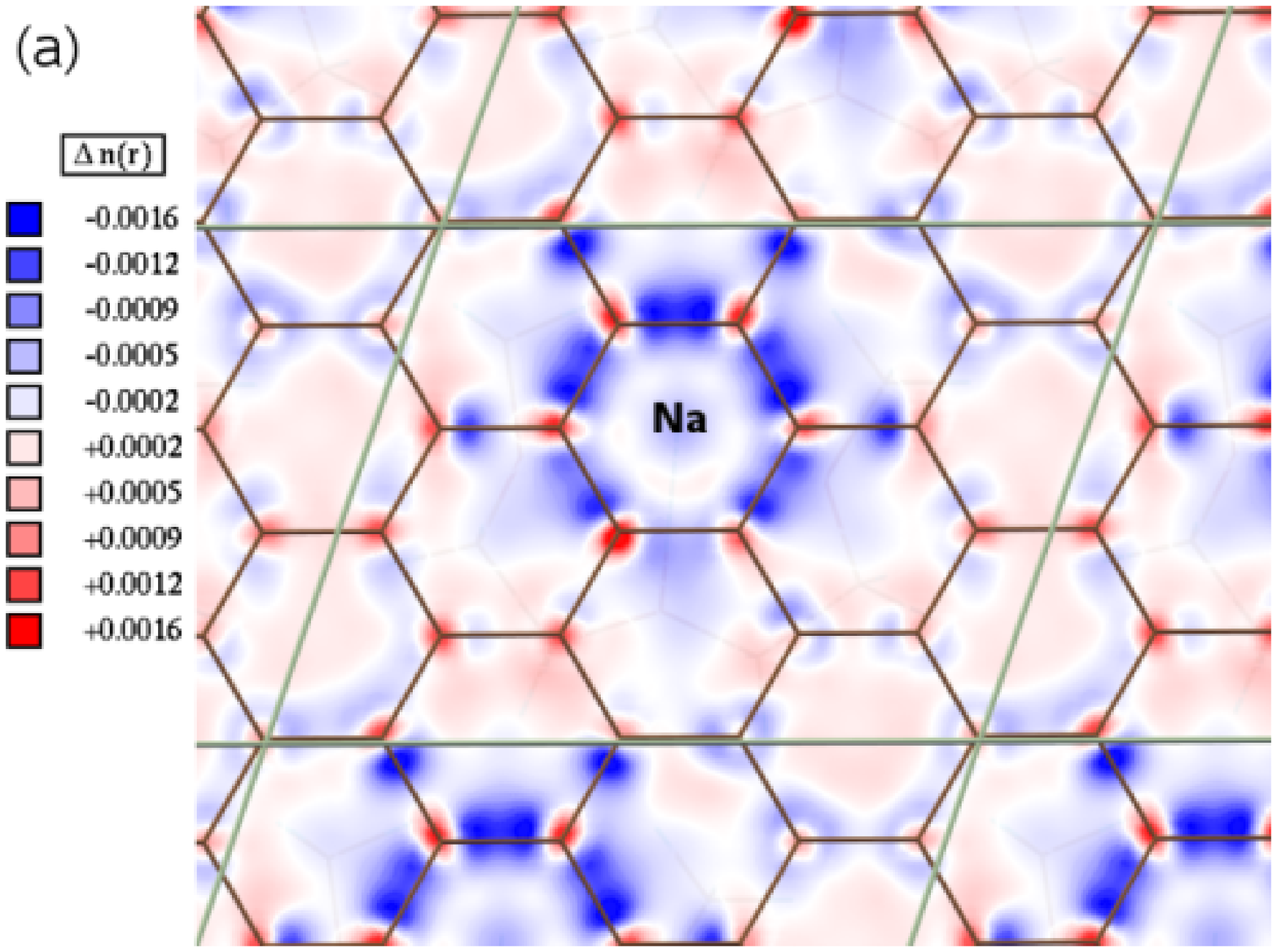} \hspace*{5pt}
\includegraphics[clip=true,scale=0.47]{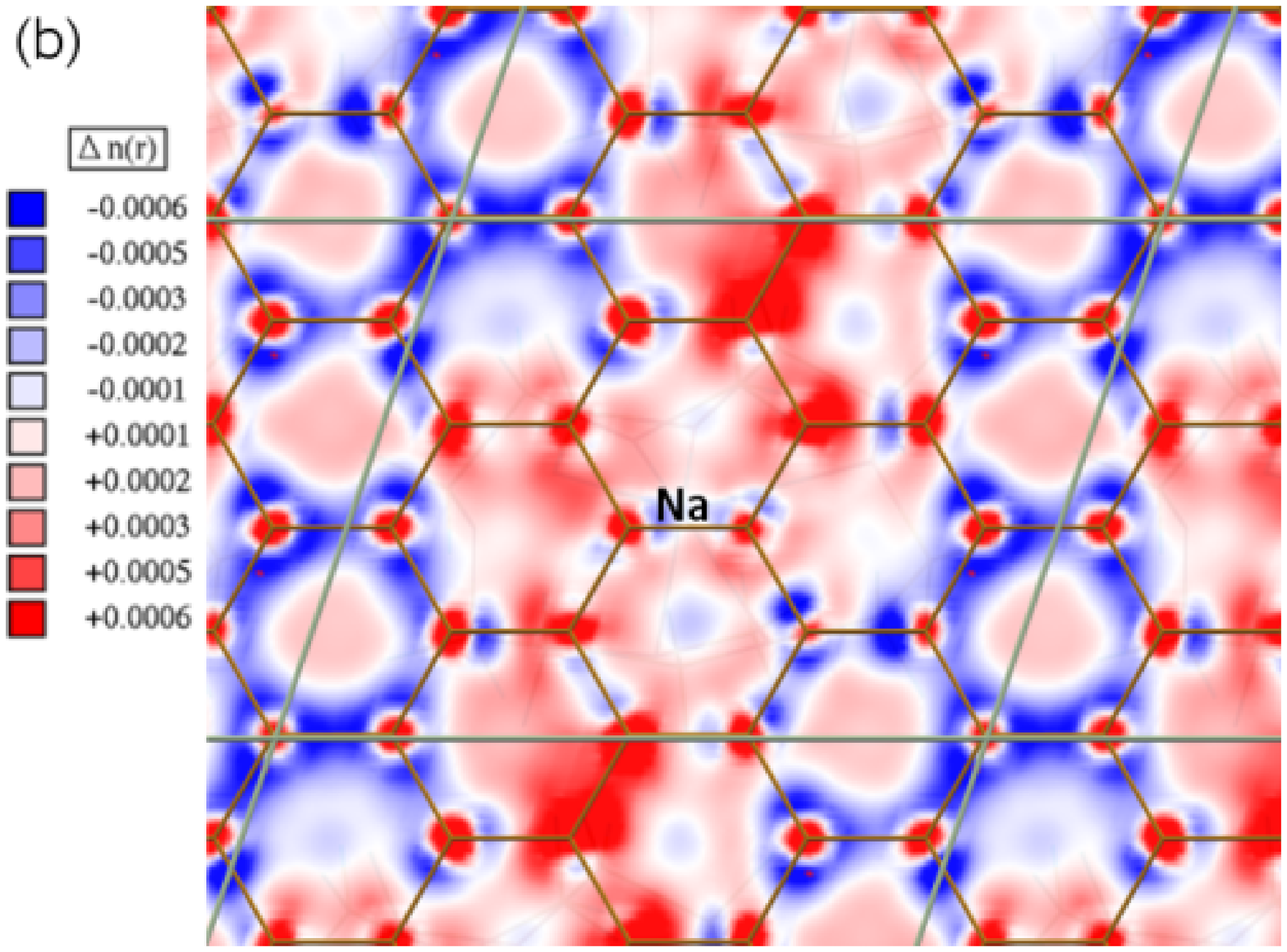} 
\end{center}
\caption{\label{fig_dens}Isoline plots of in-plane distribution of charge density difference at graphene layer in (a) Na(digl)C$_{20}$ and (b) Na(digl)$_2$C$_{20}$ compounds. Negative value indicates the donation of electron, while positive value the acceptance of electron.}
\end{figure*}

\begin{table}[!th]
\begin{center}
\caption{Atomic charge of graphene layer, diglyme molecule (C, H, O), and Na or Li atom in $t$-GICs Na(digl)C$_n$, Na(digl)$_2$C$_n$ and Li(digl)$_2$C$_n$ with $n=18$, 20, 24, 28, and 32, calculated by Mulliken population analysis method. Values in parenthesis indicate charge per atom. Positive values mean the accepted electron, while negative values the donated electron.}
\label{tab_mul}
\begin{tabular}{cccccccc}
\hline
 &          & \multicolumn{4}{c}{Diglyme} & & \\
\cline{3-6}
$n$ & Graphene & C & H & O & Total & Na / Li & Sum \\
\hline
 & \multicolumn{7}{c}{Na(digl)C$_n$} \\
18&0.194 (0.011)&$-$3.510 ($-$0.585)&3.360 (0.240)&0.092 (0.031)&$-$0.058&$-$0.142&$-$0.006 \\
20&0.170 (0.009)&$-$3.542 ($-$0.590)&3.360 (0.240)&0.116 (0.039)&$-$0.066&$-$0.108&$-$0.004 \\
24&0.232 (0.010)&$-$3.738 ($-$0.623)&3.568 (0.255)&0.114 (0.038)&$-$0.056&$-$0.174&~~0.002 \\
28&0.285 (0.010)&$-$3.536 ($-$0.589)&3.285 (0.235)&0.131 (0.044)&$-$0.120&$-$0.166&$-$0.001 \\
32&0.310 (0.010)&$-$3.998 ($-$0.666)&3.744 (0.267)&0.132 (0.044)&$-$0.122&$-$0.191&$-$0.003 \\
\hline
 & \multicolumn{7}{c}{Na(digl)$_2$C$_n$} \\
18&0.298 (0.017)&$-$7.266 ($-$0.606)&6.832 (0.244)&0.296 (0.049)&$-$0.138&$-$0.160&~~0.000 \\
20&0.368 (0.018)&$-$7.275 ($-$0.606)&6.774 (0.242)&0.290 (0.048)&$-$0.211&$-$0.152&~~0.005 \\
24&0.315 (0.013)&$-$7.558 ($-$0.630)&7.089 (0.253)&0.306 (0.051)&$-$0.163&$-$0.148&~~0.004 \\
28&0.399 (0.014)&$-$7.627 ($-$0.636)&7.092 (0.253)&0.302 (0.050)&$-$0.233&$-$0.160&~~0.006 \\
32&0.385 (0.012)&$-$8.045 ($-$0.670)&7.534 (0.269)&0.286 (0.048)&$-$0.225&$-$0.162&$-$0.002 \\
\hline
& \multicolumn{7}{c}{Li(digl)$_2$C$_n$} \\
18&0.270 (0.015)&$-$7.420 ($-$0.618)&6.809 (0.243)&0.258 (0.043)&$-$0.353&0.074&$-$0.009 \\
20&0.334 (0.017)&$-$7.427 ($-$0.619)&6.770 (0.242)&0.256 (0.043)&$-$0.401&0.062&$-$0.005 \\
24&0.278 (0.012)&$-$7.712 ($-$0.643)&7.104 (0.254)&0.266 (0.044)&$-$0.342&0.062&$-$0.002 \\
28&0.360 (0.013)&$-$7.770 ($-$0.648)&7.092 (0.253)&0.262 (0.044)&$-$0.416&0.056&~~0.000 \\
32&0.356 (0.011)&$-$8.200 ($-$0.683)&7.527 (0.269)&0.252 (0.042)&$-$0.421&0.056&$-$0.009 \\
\hline
\end{tabular}
\end{center}
\end{table}
%

%